\documentclass[11pt,a4paper]{article}
\usepackage[dvipdfmx]{graphicx}
\usepackage{bmpsize}
\usepackage{jheppub}

\usepackage{amsmath,amssymb}
\usepackage{bm}
\usepackage{color}
\usepackage{bm,float,epsfig,multirow}
\def\Babar{{\mbox{\slshape B\kern-0.1em{\smaller A}\kern-0.1em B\kern-0.1em{\smaller A\kern-0.2em R}}}}

\definecolor{mygrn}{rgb}{0,0.65,0}

\newcommand{\eq}[1]{Eq.~(\ref{#1})}
\newcommand{\eqs}[1]{Eqs.~(\ref{#1})}
\newcommand{\rf}[1]{Ref.~\cite{#1}}

\newcommand{\GkII}{I\hspace{-.1em}I}
\newcommand{\GkIII}{I\hspace{-.1em}I\hspace{-.1em}I}
\newcommand{\GkIV}{I\hspace{-.1em}V}

\newcommand{\GkVI}{V\hspace{-.1em}I}
\newcommand{\GkVII}{V\hspace{-.1em}I\hspace{-.1em}I}

\usepackage[normalem]{ulem}

\title{CP-violating phase on magnetized toroidal orbifolds}

\author[a]{Tatsuo~Kobayashi,}
\author[b,1]{Kenji~Nishiwaki,\note{Corresponding author.}}
\author[c]{Yoshiyuki~Tatsuta\,}

\affiliation[a]{Department of Physics, Hokkaido University, Sapporo 060-0810, Japan}
\affiliation[b]{School of Physics, Korea Institute for Advanced Study, Seoul 02455, Republic of Korea}
\affiliation[c]{Department of Physics, Waseda University, Tokyo 169-8555, Japan}

\emailAdd{kobayashi@particle.sci.hokudai.ac.jp}
\emailAdd{nishiken@kias.re.kr}
\emailAdd{y\_tatsuta@akane.waseda.jp}

\abstract{
We study the CP-violating phase of the quark sector %in the $U(8)$ flavor model
on $T^2/Z_N \, (N=2,3,4,6)$ with non-vanishing magnetic fluxes, where properties of possible origins of the CP violation are investigated minutely.
In this system, a non-vanishing value is mandatory in the real part of the complex modulus parameter $\tau$ of the two-dimensional torus {in order to explain the CP violation in the quark sector}.
On $T^2$ without orbifolding, underlying discrete flavor symmetries severely restrict the form of Yukawa couplings and it is very difficult to reproduce the observed pattern in the quark sector including the CP-violating phase $\delta_{\rm CP}$.
{When} multiple Higgs doublets {emerge} {on $T^2/Z_2$}, the mass matrices of the zero-mode fermions can be written in {the} Gaussian textures by choosing appropriate configurations of vacuum expectation values of the Higgs fields.
When such Gaussian textures of mass matrices are realized, we show that all of the quark profiles, which are mass hierarchies among the quarks, quark mixing angles, {and $\delta_{\rm CP}$} can be simultaneously realized.
}

\subheader{Report number: EPHOU-16-018, KIAS-P16072, WU-HEP-16-19}
\arxivnumber{1609.08608}

\allowdisplaybreaks[3]
\begin{document}

\maketitle
\flushbottom

%0
%%%%%%%%%%%%%%%%%%%%%%%%%%%%%%%%%%%%%%
\section{Introduction}
%%%%%%%%%%%%%%%%%%%%%%%%%%%%%%%%%%%%%%

The standard model (SM) of particle physics {had} been completed by the discovery of the Higgs boson~\cite{Aad:2012tfa, Chatrchyan:2012ufa}, and it is known that the SM is a quite successful theory which can explain almost all {the} phenomena around the electroweak scale with great accuracy.
However, since we know that some theoretical difficulties prevail in the SM, various phenomenological models beyond the SM of particle physics have been proposed and investigated steadily.

Addressing extra dimensions is known to be an avenue to distinctive phenomenological model buildings.
Indeed, several models can solve phenomenological problems, e.g., the gauge hierarchy problem~\cite{ArkaniHamed:1998rs,Antoniadis:1998ig,Randall:1999ee}, doublet-triplet splitting problem in supersymmetric $SU(5)$ grand unified theory (GUT)~\cite{Kawamura:2000ev, Kawamura:2000ir}, tiny neutrino masses~\cite{ArkaniHamed:1998vp}, the flavor puzzle among the SM quarks and leptons~\cite{Agashe:2004cp, Altarelli:2005yp} and so on.
On the other hand, superstring theories predict the existence of extra {six} dimensions~(6D) due to their theoretical consistency, i.e., the totally ten dimensions (10D) {are involved}.
Phenomenologies inspired by superstring theory have been enthusiastically studied from the late {1980's}.
Thus, from both of phenomenological bottom-up and theoretical top-down points of view, {models on} extra dimensions {play an important role} in concrete model building.

%{Obviously, six-dimensional models have larger degrees of freedom in boundary conditions than five dimensional models.
%On top of that, higher dimensional gauge theories such as supersymmetric Yang--Mills (SYM) theories compactified on two-dimensional (2D) torus $T^2$, or toroidal orbifolds $T^2/Z_N \, (N=2, 3,4,6)$.}
Recently, higher dimensional supersymmetric Yang--Mills (SYM) theories compactified on two-dimensional (2D) torus $T^2$, or toroidal orbifolds $T^2/Z_N \, (N=2, 3,4,6)$, {as well as 6D torus and orbifolds} have attracted much attentions.
{Obviously, six-dimensional models have larger degrees of freedom in boundary conditions than five-dimensional models. 
On top of that, higher dimensional gauge theories such as the SYM theory} are more interesting since we can put non-vanishing magnetic fluxes{, which} are vacuum expectation values of {extra-dimensional} components of a higher dimensional vector field.
Indeed, {such} higher dimensional SYM {theories} with non-vanishing magnetic fluxes are often used in the context of higher dimensional (supersymmetric) GUTs.
It is attractive to use such magnetic fluxes from the standpoint of theoretical and phenomenological model buildings, because the SYM theory compactified on 6D spacetime including 2D torus with magnetic fluxes can lead to several phenomenological ingredients, e.g., {chiral} (supermultiplet) matter fields, their generations, and (three-point) Yukawa couplings and so on~\cite{Cremades:2004wa}.
Indeed, phenomenological aspects of the higher dimensional SYM theory with non-vanishing magnetic fluxes have been investigated, e.g., extensions to toroidal orbifolds \cite{Abe:2008fi, Fujimoto:2013xha}, {analyses} of zero-mode wavefunctions \cite{Abe:2013bca, Abe:2014noa}, three-generation models \cite{Abe:2008sx, Abe:2015yva}, the minimal supersymmetric standard model (MSSM) and its modified models \cite{Abe:2012fj, Abe:2013bba}, non-Abelian discrete flavor symmetries \cite{Abe:2009vi,Abe:2009uz,Abe:2010iv,Abe:2014nla}, realistic flavor structures of quarks and leptons \cite{Abe:2014vza}, and {other studies~\cite{Abe:2014soa,Abe:2009dr,Buchmuller:2015eya,Abe:2015jqa,Buchmuller:2015jna,Buchmuller:2016dai,Buchmuller:2016bgt}}.

The four-dimensional CP symmetry can be embedded into 10D Lorentz symmetry with positive determinant
in 10D SYM theories and superstring theory~\cite{Green:1987mn,Strominger:1985it}.
That is, we can  combine the 4D CP transformation and extra 6D transformation with negative determinant 
to make a 10D proper Lorentz transformation. 
For example, when we denote the complex coordinates of the extra 6D space by $z^a$ {($z^a \equiv y_{2a+3} + i\,y_{2a+4}$, $a=1,2,3$)}, 
the 4D CP transformation with $z^a \rightarrow (z^a)^*$ is the 10D Lorentz transformation.
That is a good symmetry on the trivial background.\footnote{{That provides a 
solution of the strong CP problem~\cite{Dine:1992ya,Choi:1992xp}.}}
We may also have other embedding, whose transformation of the extra 6D space leads to 
the transformation with negative determinant.
However, {nontrivial} geometrical and gauge backgrounds violate such a CP symmetry embedded 
in higher dimensions and the 4D CP symmetry could be violated~\cite{Lim:1990bp,Kobayashi:1994ks,Lim:2009pj}.
Then, {CP-violating} phases would appear in 4D low-energy effective field theory.
For example, a certain type of orbifolds break the symmetry $z^a \rightarrow (z^a)^*$~\cite{Kobayashi:1994ks,Lim:2009pj} 
and 
magnetic fluxes also break such a symmetry.
Violation of the CP symmetry does not always lead to non-vanishing Kobayashi--Maskawa~(KM) CP phase {$\delta_{\rm CP}$}~\cite{Kobayashi:1973fv}
(at tree level), although it would lead to some CP-violating terms. 
Obviously, Yukawa couplings should have {nontrivial} CP phases, 
which cannot be canceled by rephasing fields.
In higher dimensional SYM {theories}, the Yukawa coupling is obtained by the product of 
the higher dimensional gauge coupling and overlap integral of zero-mode wavefunctions in 
the compact space.
Thus, the wavefuncitons and its overlap integral must be {nontrivial} to realize 
the physical CP phase.
The constant zero-mode profiles on the simple torus and oribifold could not lead to non-vanishing 
physical CP phase within the framework of pure SYM {theories}.
Also, the coupling selection rule as well as the flavor symmetry would be 
important to realize non-vanishing mixing and KM phase~\cite{Kobayashi:1994ks}.
 
In this paper, we study KM {CP-violating} phase within the framework of 
10D SYM {theories} on magnetized orbifolds.
Zero-mode profiles on the torus and {orbifolds} are quite {nontrivial}, 
and their couplings include {nontrivial} phases.
The models on the torus with magnetic flux has a {large} symmetry~\cite{Abe:2009vi}
{and} orbifolding would be helpful to violate such a symmetry and to realize 
non-vanishing physical CP phase.

We study the CP-violating phase of the quark sector %in the $U(8)$ flavor model
on $T^2/Z_N$ $(N=2,3,4,6)$ with non-vanishing magnetic fluxes.
In {the case} of multiple Higgs doublets emerging, the mass matrices of the zero-mode fermions can be written in {the} Gaussian {texture} by choosing appropriate configurations of vacuum expectation values of the Higgs fields \cite{Abe:2014vza}.
When such Gaussian textures of mass matrices are realized, we show that all of the quark profiles, which are mass hierarchies among the quarks, quark mixing angles including the CP-violating phase $\delta_{\rm CP}$, can be simultaneously realized.

This paper is organized as follows.
In Sec.~\ref{sec:formula}, we briefly review Yukawa couplings and mass matrices on magnetized orbifolds $T^2/Z_N$
$(N = 2,3,4,6)$.
Then, we focus on the quark sector of the SM and identify the Yukawa couplings and the mass matrices with those of the quarks.
{In Sec.~\ref{sec:example}, we study what is important to realize non-vanishing $\delta_{\rm CP}$ 
in simple examples.}
In Sec.~\ref{sec:result}, we introduce our strategy of analyzing the CP-violating phase $\delta_{\rm CP}$ in the quark sector and show results of numerical calculations with mentioning possible origins of the CP-violating phase in the quark sector on magnetized toroidal orbifolds $T^2/Z_N$.
{Section}~\ref{sec:summary} is devoted to conclusions and discussions.
{In Appendix~\ref{appendix:others}, we provide additional examples of numerical calculations under different setups on magnetized orbifolds.}

%0
%%%%%%%%%%%%%%%%%%%%%%%%%%%%%%%%%%%%%%%%%%%%%%%%%%%%%%%%%%%%%%%%%%%%%%
\section{Quark mass matrix on magnetized $T^2/Z_N$
\label{sec:formula}}
%%%%%%%%%%%%%%%%%%%%%%%%%%%%%%%%%%%%%%%%%%%%%%%%%%%%%%%%%%%%%%%%%%%%%%

In this section, we briefly review the form of the mass matrices on toroidal orbifolds $T^2/Z_N$ ($N=2,3,4$ and $6$) with magnetic fluxes {in 6D spacetime}, based on \cite{Abe:2015yva, Fujimoto:2016zjs}.
{The} 2D orbifold among the 6D space is 
important for the flavor structure.
Thus, we concentrate ourselves on the 2D orbifold part here.

First of all, we consider three sectors ({labelled as} ``1'', ``2'', ``3'') which constitute (three-point) Yukawa-type interactions.
The analytical expression of the Yukawa couplings on a magnetized $T^2/Z_N$ {is given} as
%%%
\begin{gather}
 \widetilde{\lambda}'_{I',J',K'} = \sum_{I=0}^{|M_1|-1} \sum_{J=0}^{|M_2|-1} \sum_{K=0}^{|M_3|-1}
\lambda_{I, J, K} \left({\cal U}^{Z_N; \eta_1}\right)_{I,I'} \left({\cal U}^{Z_N; {\eta_2}}\right)_{J,J'}
\left({{\cal U}^{Z_N; {\eta_3}}}\right)_{K,K'}^\ast,
\label{yukawa}
\end{gather}
%%%
where $\lambda_{I, J, K}$ represents the Yukawa couplings on a magnetized $T^2$,
%%%
\begin{gather}
\lambda_{I, J, K} = \sum_{m \in Z_{M_3}}
\delta_{I + \alpha_1 + J + \alpha_2 + m M_1, \, K + \alpha_3 + \ell M_3} \times
\vartheta \left[
\begin{array}{c}
\frac{M_2 \left(I + \alpha_1 \right) - M_1 \left(J + \alpha_2 \right) + m M_1 M_2}{M_1 M_2 M_3} \\[5pt] 0
\end{array}
\right] {(X, Y)}, \label{torusyukawa}\\
X := M_1 \beta_2 - M_2 \beta_1, \qquad Y := M_1 M_2 M_3 \tau,
\end{gather}
%%%
where $I,\,J,\,K$ identify the degenerated states on the magnetized torus of the sectors ``1'', ``2'', ``3'', respectively.
In the low energy effective theory, zero-modes of the $X$-th matter fields $(X = 1, 2, 3)$ are characterized by the magnitudes of magnetic fluxes $M_X$, Scherk--Schwarz~(SS) phases $\alpha_X$ and $\beta_X$, and $Z_N$ parities $\eta_X$, respectively.
{We note that this form is for the case of $M_{1,2,3} > 0$.}
{Here, the consistency conditions are held:}
\begin{equation}
{M_1 + M_2 = M_3,\quad
\alpha_1 + \alpha_2 = \alpha_3,\quad
\beta_1 + \beta_2 = \beta_3,\quad
\eta_1 \eta_2 (\eta_3)^\dagger = 1.}
\label{eq:consistency_condition}
\end{equation}
The complex structure modulus parameter of the torus is designated by $\tau$.
Here, we ignored an overall factor, which is a function of $M_{1,2,3}$.
$I',\, J'$ and $K'$ label the (physical) flavor eigenstates of the three kinds of matter fields {appearing} the Yukawa interaction, where the kinetic terms of the zero-modes are suitably diagonalized, whereas the Yukawa coupling matrix is not yet diagonalized.
The index $I' \, (J', K')$ runs from $0$ to ${\text{rank}} \, [{\cal K}^{(Z_N; {\eta_{1}})}] -1 \, ({\text{rank}} \, [{\cal K}^{(Z_N; {\eta_{2}})}] -1, {\text{rank}} \, [{\cal K}^{(Z_N; {\eta_{3}})}] -1)$, respectively.
$\vartheta$ denotes the Jacobi's theta function whose definition is given as
%%%
\begin{gather}
\vartheta \left[
\begin{array}{c}
 a\\ b
\end{array}
\right] (c\nu,c\tau) 
=\sum_{l=-\infty}^{\infty}e^{i\pi (a+l)^2c\tau}e^{2\pi i(a+l)(c\nu +b)},
\label{jacobi}
\end{gather}
{where $a$ and $b$ are real numbers, $c$ is an integer, and $\nu$ and $\tau$ are complex numbers with $\text{Im}\,\tau > 0$, respectively.}
%%%
In~\eq{yukawa}, it is easily found that the Yukawa couplings on $T^2$ are mixed by the matrices,
%%%
\begin{gather}
({\cal U}^{Z_N; \eta_{X}})_{\mathcal{I}, \, \mathcal{I}'} = \sum_{\mathcal{I}'' = 0}^{|M_X|-1} ({\cal K}^{(Z_N; \eta_{X})})_{\mathcal{I}, \mathcal{I}''} (U^{(Z_N; \eta_{X})})_{\mathcal{I}'', \mathcal{I}'},
\label{coefficients}
\end{gather}
%%%
where ${\cal K}^{(Z_N; \eta_{X})}$ and $U^{(Z_N; \eta_{X})}$ describe the effects via the projection from $T^{2}$ onto $T^{2}/Z_{N}$ and the diagonalization of the kinetic terms of the sector $X$, respectively{.}
{Concrete forms of ${\cal K}^{(Z_N; \eta_{X})}$ and $U^{(Z_N; \eta_{X})}$ are provided in~\rf{Fujimoto:2016zjs}}.
In the following part, we identify the first matter fields in the sector ``$1$'' with left-handed quarks, the second ones in the sector ``$2$'' with right-handed quarks and the third ones in the sector ``$3$'' with Higgs doublet fields.
Also, we require three generations in the left-handed and right-handed quarks.
Thus, in the following, we assume the {primed} indices among the left-handed and right-handed quarks run over $I', J'=0,1,2$.
We note that possibilities to realize such three-generation models {were} investigated in \rf{Abe:2013bca}.\footnote{See also \rf{Abe:2015yva}.}

In a case with ${n_H}$ generations in the Higgs fields, i.e., $K' = 0,1,...,{n_H-1}$, the quark mass matrix consists of the Yukawa couplings in Eq.~(\ref{yukawa}) and the Higgs {vacuum expectation values~(VEVs)} $(v_{K'})$ as
%%%
\begin{gather}
M_{I', J'} = \sum_{K' = 0}^{{n_H-1}} \widetilde{\lambda'}_{I', J', K'} \times  v_{K'}.
\end{gather}
%%%
%Here and hereafter we abbreviate prime symbols $(')$ in flavor indices.
In this paper, we do not construct complete models where all of {ten-dimensional~(10D)} configurations are manifested (not only the flavor part of {the quark and Higgs sectors}).
We respect a procedure of classifying possible configurations of the quark sector in~\rf{Abe:2014vza}, and consider the case of multiple up- and down-type Higgs doublets emerging {as,}
%%%
\begin{gather}
(M_u)_{I', J'} = \sum_{K' = 0}^{{n_H-1}} {\widetilde{\lambda'}_{I',J',K'}} \times v_{uK'}, 
	\label{eq:up_Yukawa_form}\\
(M_d)_{I', J'} = \sum_{K' = 0}^{{n_H-1}} {\widetilde{\lambda'}_{I',J',K'}} \times v_{dK'},
	\label{eq:down_Yukawa_form}
\end{gather}
%%%
where $v_{uK'}\, (v_{dK'})$ are VEVs of up-type (down-type) Higgs fields in the MSSM-like Higgs sector, respectively.
In the next section, we analyze the above type of mass matrices in order to investigate the CP-violating phase in the quark sector.

%0
%%%%%%%%%%%%%%%%%%%%%%%%%%%%%%%%%%%%%%%%%%%%%%%%%%%%%%%%%%%%%%%%%%%%%%
\section{Basic properties of a CP-violating phase}
\label{sec:example}
%%%%%%%%%%%%%%%%%%%%%%%%%%%%%%%%%%%%%%%%%%%%%%%%%%%%%%%%%%%%%%%%%%%%%%

{In this section, we survey basic properties of the CP-violating phase originating from the quark Yukawa couplings of the magnetized orbifold model, where the model is expected to be derived from the 10D {supersymmetric} Yang--Mills theory which is a low-energy effective theory of superstring theories.
As widely known, observing complex degrees of freedom in the Yukawa couplings (in flavor eigenstates) is a necessary condition for realizing a CP-violating phase (in quark mass eigenstates)~\cite{Kobayashi:1973fv}.

Extra-spacial components of a 10D vector field behave as scalars {from the four-dimensional point of view}, which can be candidates of (multiple) Higgs bosons.
Here, no complex parameter exists in the 10D action, but this is not the end of the story.
A CP-violating phase can be realized in the following reason.
On the magnetized $T^2$, the mode function with the flavor index $I$ of the matter field $X$ which feels the magnetized flux $M_X\,(>0)$ and SS phases ($\alpha_X$, $\beta_X$) is given as the following form
\begin{equation}
\Theta^{(I + \alpha_{X}, \beta_{X})}_{M_{X}} (z, \tau) \propto
e^{i\pi M_{X} \, {z} \, {\mathrm{Im}(z)\over \mathrm{Im}\tau}}\cdot \vartheta \left[
\begin{array}{c}
{I + \alpha_{X} \over M_{X}} \\ -\beta_{X}
\end{array}
\right] (M_{X} \, {z}, M_{X} \, \tau),
\label{eq:zero-mode_function}
\end{equation}
where an overall factor is not important in the following discussions and is neglected.\footnote{
{When $M_X < 0$, the anti-holomorphic counterpart ($z \to \bar{z}$, $\tau \to \bar{\tau}$) comes in the mode function.}}
As discussed in Refs.~\cite{Lim:1990bp,Kobayashi:1994ks,Lim:2009pj} {and we touched in Introduction}, in even {spacetime} dimensions, under the ``modified'' CP transformation, which corresponds to the ordinary ``4D'' CP transformation after the Kaluza--Klein decomposition, the complex coordinate $z$ is transformed as the complex conjugation; {``4D''} CP: $z \to z^\ast$~\cite{Lim:1990bp,Kobayashi:1994ks,Lim:2009pj}.
The zero-mode function in Eq.~(\ref{eq:zero-mode_function}) is not invariant under the transformation, which is ensured by the non-zeroness of the following variables; $\text{Re}\,\tau$, $M_X$, $\alpha_X$ and/or $\beta_X$.
Then in general, effective Yukawa couplings can be complex and the above necessary condition is fulfilled.

On the other hand, the existence of the complex degrees of freedom itself is not a sufficient condition for the emergence of a CP-violating phase.
In the present model, Yukawa couplings cannot take arbitrary components, the values of which are determined by the fundamental parameters; $\tau$, $M_X$, $\alpha_X$ and $\beta_X$.
Then, if the structure of the up- and down-{type} quark sectors are similar, the degree of freedom of a complex phase is 
{expected} to vanish because the Cabbibo--Kobayashi--Maskawa~{(CKM)} matrix $V_\text{CKM}$~{\cite{Kobayashi:1973fv}} is defined by use of the diagonalizing matrices for the sectors of the up-type quarks ($U_u$) and down-type ones ($U_d$),
\begin{equation}
V_\text{CKM} \equiv U_{u} (U_{d})^\dagger,
\end{equation}
whose physical components are usually parametrized as
%%%%
\begin{gather}
V = V_{\rm CKM} =
\begin{pmatrix}
c_{12} c_{13} &s_{12} c_{13} & s_{13} e^{-i\delta_{\rm CP}} \\
-s_{12}c_{23} - c_{12} s_{23} s_{13} e^{i\delta_{\rm CP}} & c_{12}c_{23} - s_{12}s_{23} s_{13}e^{i\delta_{\rm CP}} & s_{23}c_{13} \\
s_{12}s_{23}-c_{12}c_{23}s_{13}e^{i\delta_{\rm CP}} & -c_{12}s_{23} - s_{12}c_{23}s_{13}e^{i\delta_{\rm CP}} & c_{23} s_{13}
\end{pmatrix}.
\end{gather}
%%%%
It is noted that if the structures of the {up- and down-type quarks} are completely the same, the relation $U_u = U_d$ results in the trivial mixing $V_{\text{CKM}} \to \mathbf{1}_3$.
}

In {the following part}, we show illustrating samples for the {non-vanishing} CP violation where the realization of the flavor structures, e.g., the quark mass hierarchies and mixing angles are not seriously considered {at this stage.}
{We} focus on the value of the CP-violating phase {$\delta_{\rm CP}$}.
By showing such illustrating samples, we try to {exemplify} several origins of the CP-violating phases on magnetized orbifolds.
{Before going to realizations of realistic flavor structures, it would be very important to become familiar with basic characteristics of the CP violation in this system.}
%Here, we do not take care the flavor structures, e.g., the quark mass hierarchies and mixing angles, and we focus only on what can be the origins of the CP-violating phase.
{For our purpose, we utilize} the Jarlskog invariant $J_{\rm CP}$ \cite{Jarlskog:1985ht,Jarlskog:1985cw}, which is defined as
\begin{gather}
J_{\rm CP} = {\rm Im} \, [V_{12} V_{22} V_{12}^* V_{21}^*]{.}
\end{gather}
%where $V_{ij}$ is the quark mixing matrix, called Cabbibo-Kobayashi-Maskawa matrix,
%\begin{gather}
%V = V_{\rm CKM} =
%\begin{pmatrix}
%c_{12} c_{13} &s_{12} c_{13} & s_{13} e^{-i\delta_{\rm CP}} \\
%-s_{12}c_{23} - c_{12} s_{23} s_{13} e^{i\delta_{\rm CP}} & c_{12}c_{23} - s_{12}s_{23} s_{13}e^{i\delta_{\rm CP}} & s_{23}c_{13} \\
%s_{12}s_{23}-c_{12}c_{23}s_{13}e^{i\delta_{\rm CP}} & -c_{12}s_{23} - s_{12}c_{23}s_{13}e^{i\delta_{\rm CP}} & c_{23} s_{13}
%\end{pmatrix}.
%\end{gather}
{It} is invariant under {field rephasing} operations, and then a nonzero $J_{\rm CP}$ indicates the existence of a nonzero CP-violating phase $\delta_{\rm CP}$.

{To declare the structure of the CP violation carefully, let us consider cases with one up-type Higgs and one down-type Higgs ($n_H = 1$).}
{At first, we comment on the configurations on $T^2$ without orbifolding 
to compare it with the following orbifold models.}
Here, the {magnitudes} of the magnetic fluxes are uniquely fixed for realizing {three generations} in the quarks as {$M_1 \,(=M_{1'}) = -3$, $M_2 \,(=M_{2'}) = -3$, $M_3 \,(=M_{3'}) = +6$},
where six pairs of up- and down-type Higgs boson doublets emerge.
{In the following discussion, for clarity, we adopt the notation on variables as $M_1 + M_2 + M_3 = 0$, $\alpha_1 + \alpha_2 + \alpha_3 = 0$, $\beta_1 + \beta_2 + \beta_3 = 0$, $\eta_1 \eta_2 \eta_3 = 1$, which are different from those in Eq.~(\ref{eq:consistency_condition}).}

Unfortunately on the simplest geometry, Yukawa couplings in Eq.~(\ref{torusyukawa}) obey discrete flavor symmetries and major parts of {the} matrix elements are forced to be zero.
On the present magnetic fluxes, the hidden symmetry becomes $\Delta(27)$~\cite{Abe:2009vi} and only the diagonal form and permuted ones of it are possible in the three-by-three Yukawa matrix of {Eq.}~(\ref{torusyukawa}) with a fixed $K$ (now, $K=0,1,2,3,4,5$).
Thereby, at least in the case that only one up-type Higgs and one down-type Higgs contain nonzero VEVs, there exists at least one vanishing mixing angle inevitably, and eventually and manifestly $J_{\text{CP}}$ becomes zero.\footnote{
{In {Ref.}~\cite{Kobayashi:1994ks}, it was found that the coupling selection rules are quite strong and 
we {cannot} realize non-vanishing CP phase in certain heterotic orbifold models, although 
Yukawa couplings in general have {nontrivial} CP phases.}}
Then, to violate the discrete flavor symmetry would be required for (quasi-)realistic models.
A possible way of breaking the symmetry is to introduce orbifolding.
As we concretely look in the subsequent discussion, even under the simplest $Z_2$ case, such a discrete symmetry is violated and nonzero $J_{\mathrm{CP}}$ gets to be achievable.

Next, we move to the configurations {shown in Table~\ref{samples}, where we fix the parameters for the left-handed quarks}.
In Sample I, the up-type and down-type mass matrices of the quarks originate from the same magnetic fluxes and SS twist phases.
In comparison with Sample I and {\GkII}, we {can} find whether the difference of the SS twists could be important for the CP violation or not.
Similarly, in comparison with Sample I and {\GkIII}, we can investigate the importance of the different fluxes in the up-type and down-type quarks.
In Sample {\GkIV}, we consider the different magnetic fluxes and SS twist phases.
In the following part of this section, we set ${\rm Re} \, \tau \in [-\pi, \pi]$ and ${\rm Im}\, \tau=1$, for simplicity.

\begin{table}[H]
\centering
\begin{tabular}{|c|ccc|} \hline
&  & right-handed up-types & up-type Higgs doublet \\
& left-handed quarks & $\{M_2, \alpha_2, \beta_2, \eta_2\}$ & $\{M_3, \alpha_3, \beta_3, \eta_3\}$ \\
& $\{M_1, \alpha_1, \beta_1, \eta_1\}$ & right-handed down-types & down-type Higgs doublet \\
 & & $\{M_{2'}, \alpha_{2'}, \beta_{2'}, \eta_{2'}\}$ & $\{M_{3'}, \alpha_{3'}, \beta_{3'}, \eta_{3'}\}$ \\ \hline
\multirow{2}{*}{Sample I} & \multirow{2}{*}{$\{-5, 0, 0, +1\}$} & $\{+6, 0, \tfrac12, +1\}$ & $\{-1, 0, \tfrac12, +1\}$ \\
& & $\{+6, 0, \tfrac12, +1\}$ & $\{-1, 0, \tfrac12, +1\}$ \\ \hline
\multirow{2}{*}{Sample {\GkII}} & \multirow{2}{*}{$\{-5, 0, 0, +1\}$} & $\{+6, 0, \tfrac12, +1\}$ & $\{-1, 0, \tfrac12, +1\}$ \\
& & $\{+6, \tfrac12, \tfrac12, -1\}$ & $\{-1, \tfrac12, \tfrac12, -1\}$ \\ \hline
\multirow{2}{*}{Sample {\GkIII}} & \multirow{2}{*}{$\{-5, 0, 0, +1\}$} & $\{+6, 0, \tfrac12, +1\}$ & $\{-1, 0, \tfrac12, +1\}$ \\
& & $\{+7, 0, \tfrac12, -1\}$ & $\{-2, 0, \tfrac12, -1\}$ \\ \hline
\multirow{2}{*}{Sample {\GkIV}} & \multirow{2}{*}{$\{-5, 0, 0, +1\}$} & $\{+6, 0, \tfrac12, +1\}$ & $\{-1, 0, \tfrac12, +1\}$ \\
& & $\{+7, \tfrac12, \tfrac12, +1\}$ & $\{-2, \tfrac12, \tfrac12, +1\}$ \\ \hline
\end{tabular}
\caption{Sample patterns in proving an origin of the CP violation.}
\label{samples}
\end{table}

At first, we comment on Sample I.
In Sample I, it is obviously found that the CKM matrix is {the unit matrix (irrespective of the value of $\tau$)} because the mass matrices of the up- and down-sectors are equivalent.
Then, the Jarlskog invariant {$J_{\rm CP}$ becomes zero}, and the CP symmetry still remains {in the 
mass matrices at this level}.
{The important lesson from this example is that at least either of the magnetic fluxes or the pairs of the SS phases should be different for realizing a nonzero $\delta_{\rm CP}$.}

\begin{figure}[H]
\centering
\begin{minipage}{0.55\hsize}
\centering
\includegraphics[clip, width=0.9\hsize]{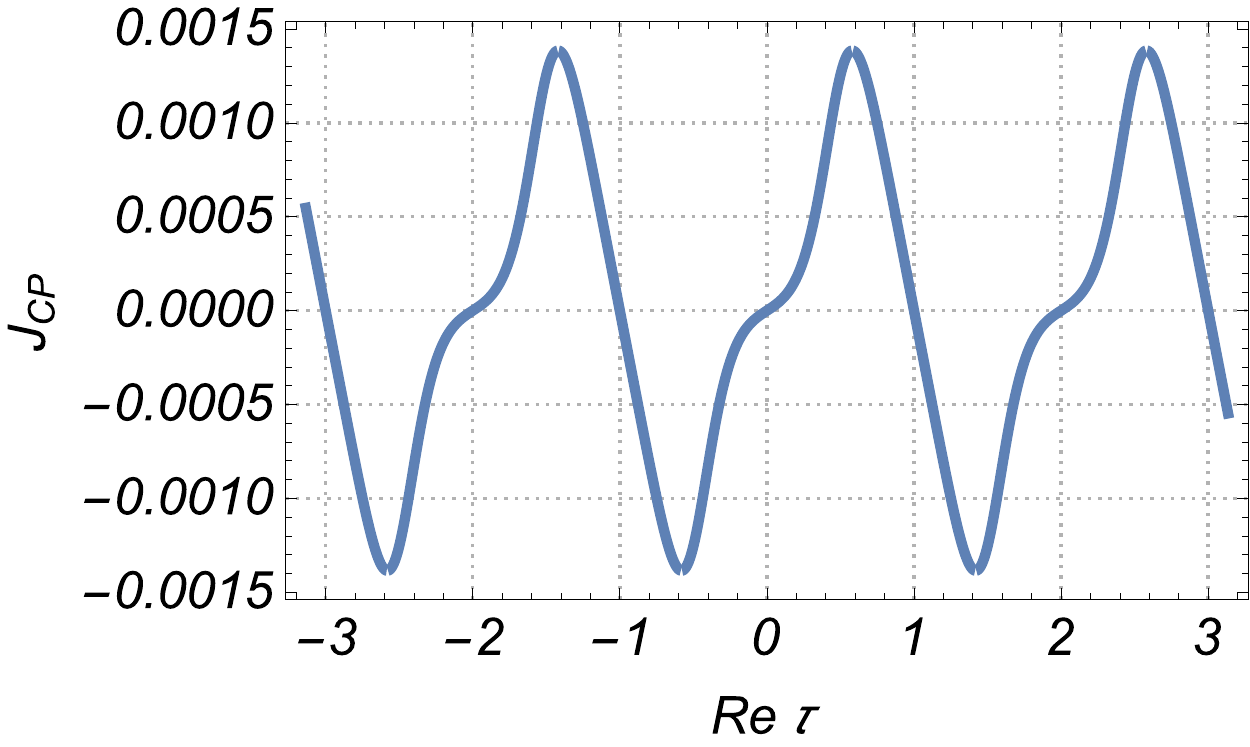}
\end{minipage}
\caption{The relationship between a real part of the complex structure modulus ${\rm Re}\, \tau$ and the Jarlskog invariant $J_{\rm CP}$ in Sample {\GkII}.}
\label{sample2}
\end{figure}

\begin{figure}[H]
\centering
\begin{minipage}{0.55\hsize}
\centering
\includegraphics[clip, width=0.9\hsize]{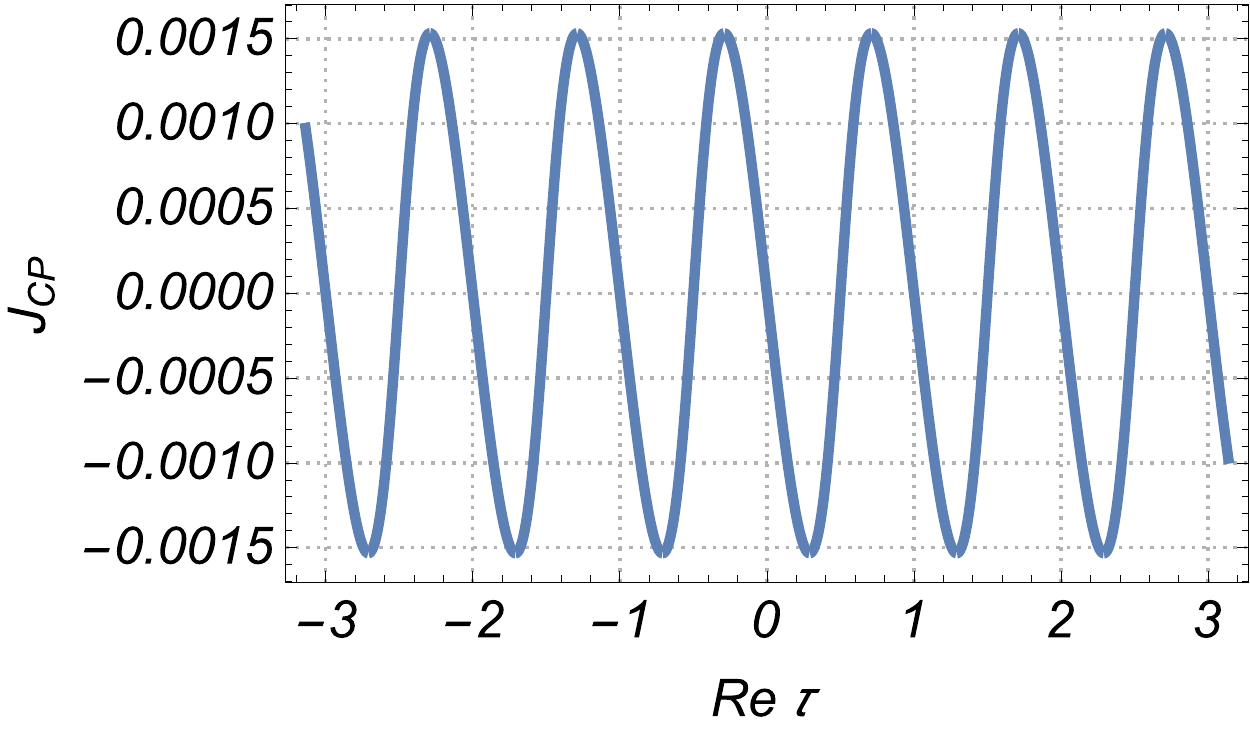}
\end{minipage}
\caption{The relationship between a real part of the complex structure modulus ${\rm Re}\, \tau$ and the Jarlskog invariant $J_{\rm CP}$ in Sample {\GkIII}.}
\label{sample3}
\end{figure}

\begin{figure}[H]
\centering
\begin{minipage}{0.55\hsize}
\centering
\includegraphics[clip, width=0.9\hsize]{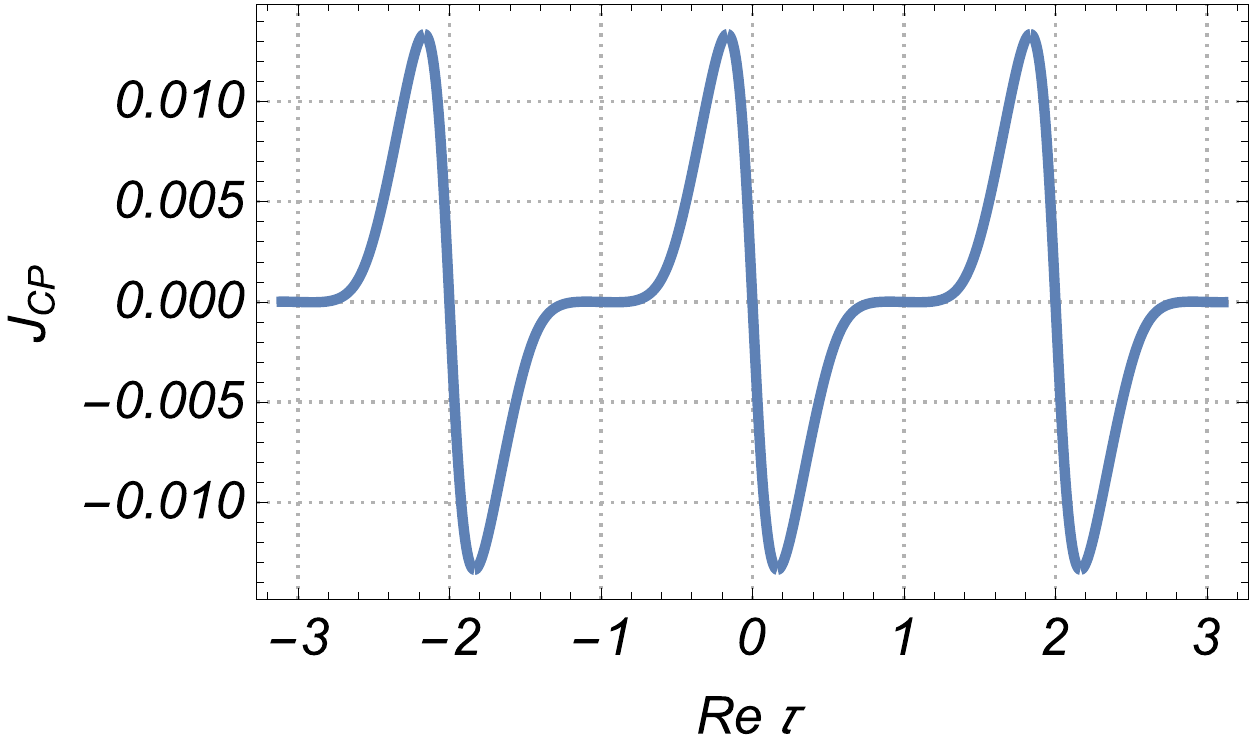}
\end{minipage}
\caption{The relationship between a real part of the complex structure modulus ${\rm Re}\, \tau$ and the Jarlskog invariant $J_{\rm CP}$ in Sample {\GkIV}.}
\label{sample4}
\end{figure}

Figures~\ref{sample2}, \ref{sample3}, \ref{sample4} show the relationships between {the} real part of the complex structure modulus ${\rm Re}\, \tau$ and the Jarlskog invariant $J_{\rm CP}$ {in the cases of \GkII, \GkIII, \GkIV, respectively}.
Figure~\ref{sample2} tells {us} that the non-integer values of ${\rm Re} \, \tau \, ({\rm Re}\, {\tau} \notin \mathbb{Z})$ break the CP symmetry.
In particular, it should be noted that the vanishing ${\rm Re} \, \tau \, ({\rm Re} \, \tau=0)$ preserve the CP 
{in the mass matrices at the tree level}, as we will see also in the other samples.
%In {figures}~\ref{sample3} and \ref{sample4}, the relationship between ${\rm Re}\, \tau$ and $J_{\rm CP}$ are shown.
Figures~\ref{sample3} and \ref{sample4} suggest us that the vanishing ${\rm Re} \, \tau \, ({\rm Re} \, \tau=0)$ does not lead to the CP breaking.
{Now, the following statement is confirmed:}
it is impossible to violate the CP without {non-vanishing ${\rm Re}\, \tau$ in the present system}.\footnote{{
In Ref.~\cite{Kobayashi:2003gf}, it was discussed that only the discrete Wilson lines, i.e., SS phases {cannot} lead to non-vanishing CP phase in heterotic orbifold models.}}

In addition, it is interesting to comment on periodic properties of ${\rm Re}\, \tau$ {in} $J_{\rm CP}$ {shown} in {Figures}~\ref{sample2}, \ref{sample3} and \ref{sample4}.
{When different patterns on SS phases are imposed in the two types of quarks, the period of $\text{Re}\, \tau$ becomes greater than one.}
{Also, {Figure}~\ref{sample4} would imply that $J_{\rm CP}$ can be amplified when differences are found both in magnetic fluxes and SS phases.}
%This implies that the SS twist phases change the magnitude of the CP violation and the correspondence between ${\rm Re}\, \tau$ and $J_{\rm CP}$.
By summarizing the results of the four illustrating samples, we can conclude that the non-vanishing ${\rm Re} \, \tau$ leads to the CP violation, and the observed values in {$J_{\rm CP} \, (\sim 10^{-5})$} would be realized by choosing the magnitude of ${\rm Re} \, \tau$, {magnetic fluxes} and the SS twist phases {suitably}.

{Finally, let us emphasize that all the observed properties are in the case with one Higgs pair ($n_H = 1$).
As described in Eqs.~(\ref{eq:up_Yukawa_form}) and (\ref{eq:down_Yukawa_form}), when multiple Higgs pairs are imposed ($n_H \geq 2$), resultant Yukawa couplings {become} superpositions of all the contributions via the multiple Higgs {boson doublets}, where correspondences between $\text{Re}\,\tau$ and $J_{\rm CP}$ would be more complicated.}

{We have shown that one can realize the {non-vanishing} CP phase in 
magnetized orbifold models by using simple examples.
However, the above models with one pair of Higgs fields {cannot} completely 
realize realistic mass ratios and mixing angles {\cite{Abe:2015yva, Fujimoto:2016zjs}}.
In the next section, we study numerically the models with multi-pairs of Higgs fields 
in order to show the possibility for realizing the quark mass ratios, 
mixing angles and the KM phase {$\delta_{\rm CP}$}.}

%0
%%%%%%%%%%%%%%%%%%%%%%%%%%%%%%%%%%%%%%%%%%%%%%%%%%%%%%%%%%%%%%%%%%%%%%
\section{Numerical analyses in Gaussian Froggatt--Nielsen models}
\label{sec:result}
%%%%%%%%%%%%%%%%%%%%%%%%%%%%%%%%%%%%%%%%%%%%%%%%%%%%%%%%%%%%%%%%%%%%%%

In this section, we numerically analyze the CP-violating phase {$\delta_{\rm CP}$} in the configuration with multiple Higgs doublet fields.
When the Higgs fields are degenerated due to non-vanishing magnetic fluxes and two or three of them contain specific non-zero VEVs, the mass matrix is approximately written in the Gaussian form \cite{Abe:2014vza} as
\begin{gather}
M_{I', J'} \sim e^{c(a_{I'} + b_{J'})^2 i \tau} = e^{-c(a_{I'} + b_{J'})^2 (\text{Im}\,\tau - i \text{Re}\,\tau)},
\end{gather}
where {$a_{I'}$, $b_{J'}$ and $c$} are {symbolized} factors which are determined by the magnitudes of the magnetic fluxes, the SS phases and the boundary conditions.
This property of the mass matrix is called {the} {\em Gaussian Froggatt--Nielsen (FN) mechanism}~\cite{Abe:2014vza}, which is a suitable mass matrix for explaining the observed flavor patterns in the quarks and charged leptons.
Here, we focus only on $N=2 $ case ($T^2/Z_2$), because the Gaussian FN mechanism is not beneficial in the cases of $N=3,4$ and $6$.
%This is easily understood as follows.
%In the $N=3,\,4$ and $6$ cases, the value of the complex structure moduli should be $\tau=e^{2\pi i /N}$.
As pointed out {in~\cite{Fujimoto:2016zjs}}, the observed mass hierarchy between the top and up quarks, i.e., $m_u/m_t = {\cal O}(10^{-5})$ cannot be realized since nontrivial large mixings {shown by} $U^{(Z_{N}; \eta_{X})}$ smear great differences originating from quasi-localized profiles of the particles due to the existence of the magnetic fluxes.
%For that reason, we focus only on the case that the Gaussian FN mass matrix is valid.
%%
Furthermore, the value of the complex modulus parameter $\tau$ is obligated to be $e^{2\pi i /N}$ for $T^2/Z_N \, (N=3,\,4,\,6)$ due to consistencies of the orbifold identifications.
{When the imaginary part of $\tau$ is less than or equal to one in the $T^2/Z_N \, (N=3,\,4,\,6)$ cases, the Gaussian FN texture does not generate sizable hierarchies in $\lambda_{I',J',K'}$ in Eq.~(\ref{yukawa}).}
%Here, the imaginary part of it is less than or equal to one, and then these cases are not suitable for the Gaussian FN mechanism.
%The FN parameter $e^{-{\rm Im} \, \tau}$ is not small, and the $T^2/Z_N \, (N=3,\,4,\,6)$ cases are not suitable for describing the observed configurations of the quarks.
Thereby, we investigate only the $T^2/Z_2$ case, where the complex modulus parameter can take an arbitrary value.
{Before showing concrete theoretical samples of quark flavor structures, we would like to comment on the moduli stabilization.
On $T^2/Z_2$ orbifold, the complex structure modulus is not geometrically stabilized.
In this paper, we assume the moduli stabilization of the complex structure modulus $\tau$, where we regard it as a free parameter.
}

Based on the previous analysis \cite{Abe:2014vza}, it is interesting to investigate a flux configuration {with five Higgs pairs $(n_H = 5)$},\footnote{The other example is shown in Appendix A.}
%%%
\begin{equation}
\begin{array}{rrr}
   M_1 = -5,&     M_2 = -7,&    M_3 = +12, \\
\eta_1 = +1,&  \eta_2 = -1,& \eta_3 = -1,
\label{M_and_eta_configuration}
\end{array}
\end{equation}
%%%
where various patterns are possible in the SS phases $\alpha_X$ and $\beta_X \, (X=1,2,3)$.
In \rf{Abe:2014vza}, {only the case with trivial SS phases $(\alpha_1, \alpha_2, \alpha_3)=(0,0,0)$ and $(\beta_1, \beta_2, \beta_3)=(0,0,0)$ were analyzed, and no discussion was made on the CP-violating phase in the quark mass matrix.}
In this paper, we take account of additional three patterns with the SS phases as shown in {Table}~\ref{fluxtable}.
Note that the difference between Pattern {I, \GkII, \GkIII, \GkIV} are found only in the values of the SS phases.
In all of the cases, the values of the Higgs VEVs are randomly selected from designated domains in our parameter searches.

\begin{table}[H]
\centering
\begin{tabular}{|c|ccc|} \hline
& left-handed quarks & right-handed quarks & Higgs fields \\
 & $\{M_1, \alpha_1, \beta_1, \eta_1\}$ & $\{M_2, \alpha_2, \beta_2, \eta_2\}$ & $\{M_3, \alpha_3, \beta_3, \eta_3\}$ \\ \hline
Pattern I & $\{-5, 0, 0, +1\}$ & $\{-7, 0, 0, -1\}$ & $\{+12, 0, 0, -1\}$ \\
Pattern {\GkII} & $\{-5, 1/2, 0, +1\}$ & $\{-7, 1/2, 0, -1\}$ & $\{+12, 0, 0, -1\}$ \\ 
Pattern {\GkIII} & $\{-5, 0, 1/2, +1\}$ & $\{-7, 0, 1/2, -1\}$ & $\{+12, 0, 0, -1\}$ \\
Pattern {\GkIV} & $\{-5, 1/2, 1/2, +1\}$ & $\{-7, 1/2, 1/2, -1\}$ & $\{+12, 0, 0, -1\}$ \\  \hline
\end{tabular}
\caption{Sample patterns of configurations. Note that Pattern I is exactly the same as that of the previous paper \cite{Abe:2014vza}.}
\label{fluxtable}
\end{table}

Here, it is important to {remind} what can be an origin of the CP-violating phase.
In order to obtain complex values in entries of Yukawa couplings~\eqref{torusyukawa} on $T^2$, we need at least non-vanishing values {in ${\rm Re} \, \tau$}.\footnote{
{In the case of $T^{2}/Z_{2}$, the form of ${\cal K}^{(Z_2; \eta)}$ is given as
\begin{align*}
({\cal K}^{(Z_2; \eta)})_{K,J} = \frac{1}{2} \left( \delta_{J,K} + \eta \, e^{-2\pi i \cdot \frac{2\beta}{M}(J+\alpha)} \delta_{-2\alpha-J,K}  \right),
\end{align*}
where $M$, $J$ and $K$, $\alpha$ and $\beta$, $\eta$ are corresponding magnetic flux, torus indices, SS phases, $Z_2$ parity, respectively~{\cite{Abe:2014noa}}.
$U^{(Z_2; \eta)}$ is easily evaluated as the unitary matrix which diagonalizes the above matrix ${\cal K}^{(Z_2; \eta)}$.
{We can find that the non-vanishing SS phases $\alpha$ and $\beta$ provide a complex phase in the matrix ${\cal K}^{(Z_2; \eta)}$.}}
}
On the other hand, in order to obtain complex values {in} coefficient matrices in {\eq{torusyukawa}} and \eq{coefficients}, both of $\alpha_X$ and {$\beta_{X}$} should be non-vanishing.
This implies that we need to input a non-vanishing value of ${\rm Re}\, \tau$ in Pattern I.
In the other patterns, it is more general to additionally turn on a non-vanishing value of ${\rm Re}\, \tau$ in addition to the {SS} phases {$\alpha_{X}$} and {$\beta_{X}$}.
That is, ${\rm Re} \, \tau$ would be required to fit an observed value of the CP-violating phase \cite{Bona:2005vz},
\begin{gather}
\delta_{\rm CP} = 1.208 \quad {\rm [rad]}.
\label{deltacp}
\end{gather}
In the following {enumeration}, we show numerical results of analyzing the mass matrices via Gaussian FN mechanism.

We mention input parameters and their ranges in our setups which are used to fit mass hierarchies and mixing angles among the up- and down-type quarks.
In the following numerical {analyses}, we choose {several} sample values of a complex structure modulus $\tau$ in {the} ranges of ${\rm Re}\, \tau$ {(from $-\pi$ to $\pi$ with each $\pi/10$ steps)} and ${\rm Im} \, \tau \in [1.5, 2.0]$, and randomly scattered values in the (up- and down-type) Higgs VEVs with multiple generations from the ranges as the shape of the ratios,
\begin{gather}
\rho_u \equiv \frac{v_{u4}}{v_{u1}} \in [0, 0.4], \qquad \rho_d \equiv \frac{v_{d4}}{v_{d1}} \in [0, 0.5], \qquad \rho'_d \equiv \frac{v_{d2}}{v_{d1}} \in [0, 0.01],
\label{rhos}
\end{gather}
where the others are set to be zero, {as taken in Ref.~\cite{Abe:2014vza}.} 
We examine the ratios of quark mass eigenvalues (not the absolute magnitudes of them), and then to assign the ratios is enough for our purpose. 
In the following analyses, {we also assume that the electroweak symmetry breaking is appropriately caused by a linear combination of multiple Higgs VEVs, and that other linear combinations are heavy enough.}
Here, the non-zero VEVs are chosen in order to obtain the Gaussian textures of mass matrices \cite{Abe:2014vza}.
We use relatively crude cuts by the observed data \cite{Bona:2005vz} for the ratios of the mass hierarchies in the up-quark sector,
\begin{gather}
\frac13 \leq \left(\frac{m_u}{m_t}\right)_{\rm obs.} \bigg/ \left(\frac{m_u}{m_t}\right)_{\rm theor.} \leq 3, \qquad \frac13 \leq \left(\frac{m_c}{m_t}\right)_{\rm obs.} \bigg/ \left(\frac{m_c}{m_t}\right)_{\rm theor.} \leq 3,
\label{upmasscut}
\end{gather}
also in the down-quark sector,
\begin{gather}
\frac13 \leq \left(\frac{m_d}{m_b}\right)_{\rm obs.} \bigg/ \left(\frac{m_d}{m_b}\right)_{\rm theor.} \leq 3, \qquad \frac13 \leq \left(\frac{m_s}{m_b}\right)_{\rm obs.} \bigg/ \left(\frac{m_s}{m_b}\right)_{\rm theor.} \leq 3,
\label{downmasscut}
\end{gather}
and also in the quark mixing angles,
\begin{gather}
0.20 \leq \sin \theta_{12} \leq 0.25, \qquad 0.01 \leq \sin\theta_{23} \leq 0.10.
\label{mixcut}
\end{gather}
Concrete values of the observed masses and mixing angles in the quark sector are summarized in Table~\ref{goodeg} {in Pattern $\text{\GkIV}$ as an example}.
{Table}~\ref{goodeg} shows that the quark mass hierarchies, small mixing angles and the CP-violating phase can be almost explained simultaneously on the magnetize orbifold $T^2/Z_2$ up to slight deviations from the observed values especially in $m_c/m_t$ and $\sin\theta_{13}$.

\begin{table}[H]
\centering
\begin{tabular}{|c|cc|} \hline
& Sample values & Observed values\\ \hline
$(m_u, m_c)/m_t$ & $(1.9 \times 10^{-5}, 1.9 \times 10^{-2})$ & $(1.5 \times 10^{-5}, 7.5 \times 10^{-3})$ \\
$(m_d, m_s)/m_b$ & $(6.6 \times 10^{-4}, 2.4 \times 10^{-2})$ & $(1.2 \times 10^{-3}, 2.3 \times 10^{-2})$ \\ 
$\sin \theta_{12}$ & $0.23$ & $0.2254$ \\
$\sin\theta_{23}$ & $0.075$ & $0.04207$ \\
$\sin\theta_{13}$ & $0.0085$ & $0.00364$ \\
$\delta_{\rm CP}$ & $1.2$ & $1.208$ \\ \hline
\end{tabular}
\caption{Sample values of an illustrating parameter configuration in Pattern $\text{\GkIV}$ {as an example}. Observed values are also quoted from \rf{Bona:2005vz}. Input parameters are chosen as ${\rm Re}\, \tau=0.5$, ${\rm Im}\, \tau=1.7$, $\rho_u=0.26$, $\rho_d=0.19$ and $\rho'_d=0.012$.}
\label{goodeg}
\end{table}

In numerical {analyses}, {we require the three-type conditions simultaneously, which are} on the mass ratios in the  up-type quarks~[\eq{upmasscut}], on the mass ratios in the down-type quarks~[\eq{downmasscut}], and on the two of mixing angles~[\eq{mixcut}].
The number of trials in randomly scattering parameters is $5 \times 10^4$ for each fixed value of $({\rm Re}\, \tau, {\rm Im}\, \tau)$.
When a configuration of Eq.~\eqref{rhos} passes the above cuts, we calculate the corresponding value of the CP-violating phase.
%based on inputted $({\rm Re}\, \tau, {\rm Im}\, \tau)$ and Higgs VEVs that can lead to realistic flavor structures, i.e., mass hierarchies and mixing angles,

\bigskip

\begin{figure}[H]
\begin{minipage}{0.5\hsize}
\centering
\includegraphics[clip, width=0.85\hsize]{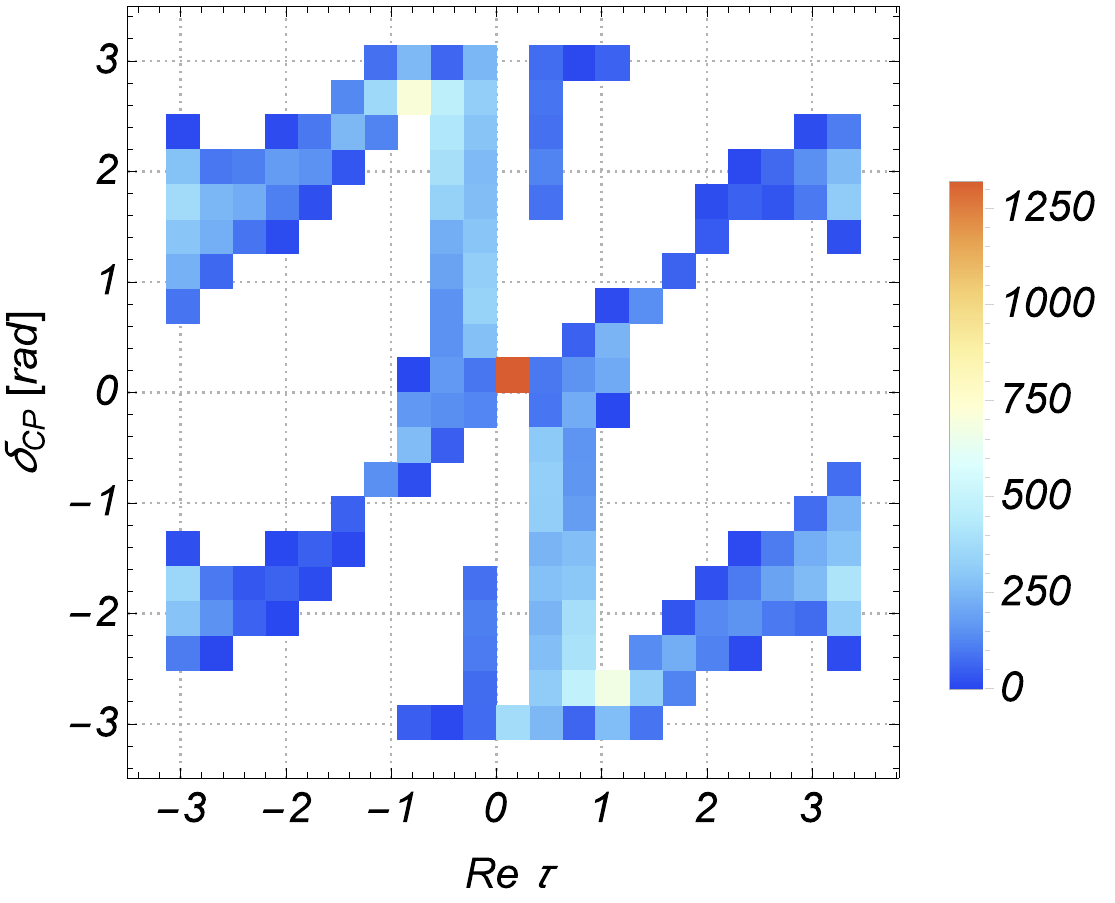}
\end{minipage}
\begin{minipage}{0.5\hsize}
\centering
\includegraphics[clip, width=0.85\hsize]{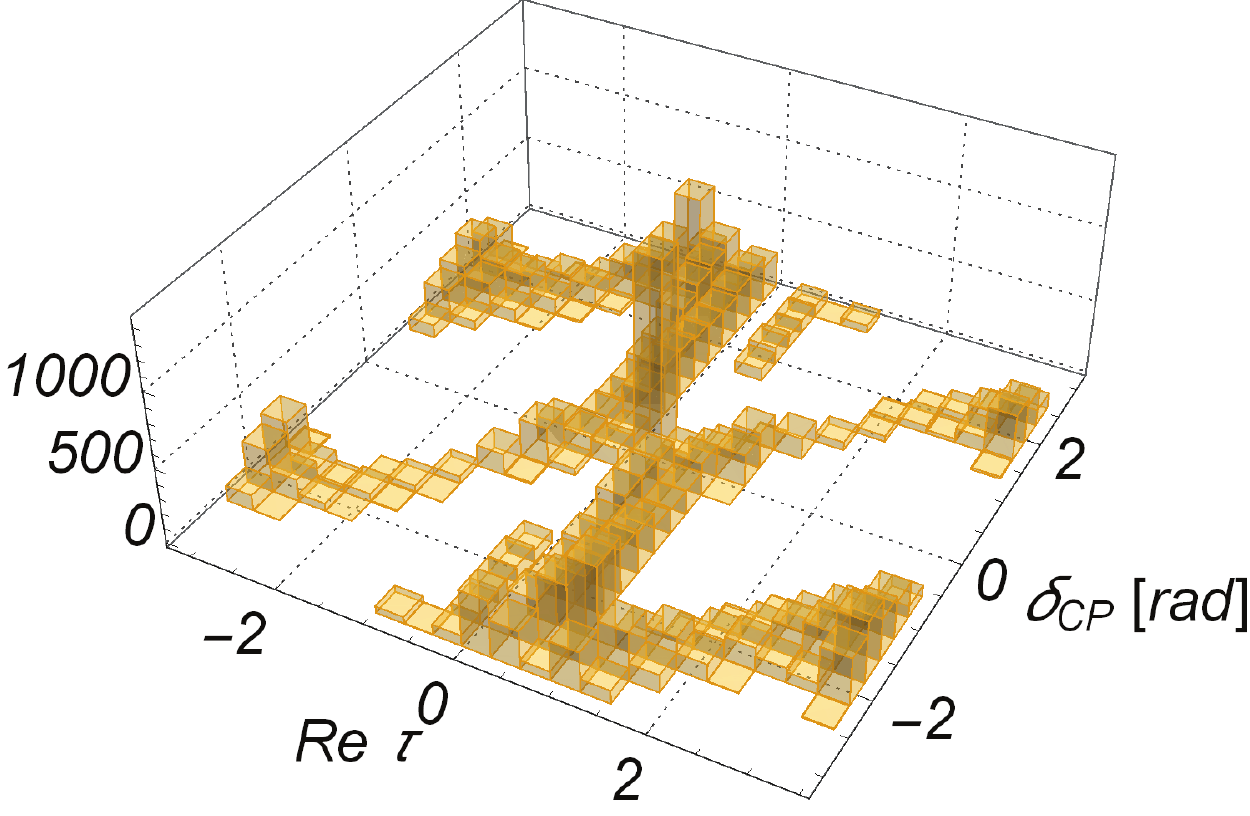}
\end{minipage}
\caption{Distributions of the CP-violating phase $\delta_{\rm CP}$ vs. real part of complex structure modulus ${\rm Re}\, \tau$ for ${\rm Im} \, \tau=1.8$ in Pattern I.}
\label{pattern1_1}
\end{figure}

\begin{figure}[H]
\begin{minipage}{0.5\hsize}
\centering
\includegraphics[clip, width=0.85\hsize]{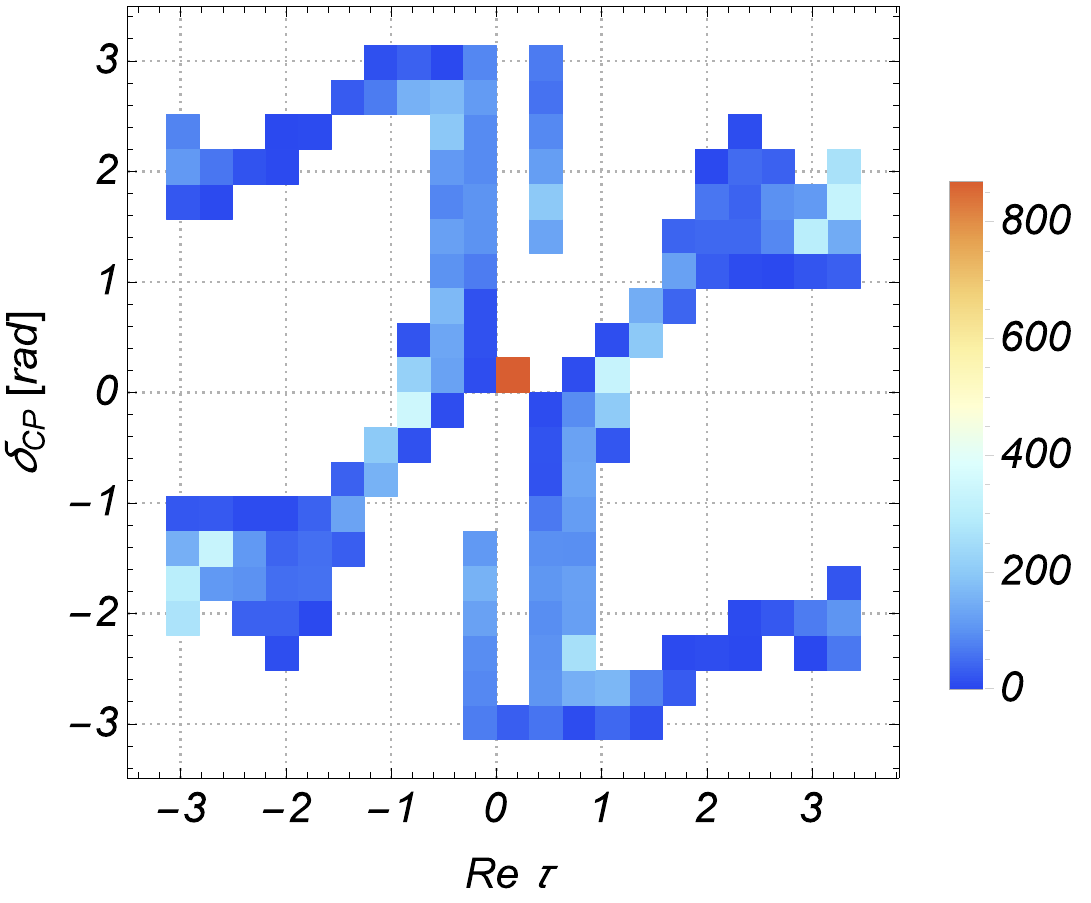}
\end{minipage}
\begin{minipage}{0.5\hsize}
\centering
\includegraphics[clip, width=0.85\hsize]{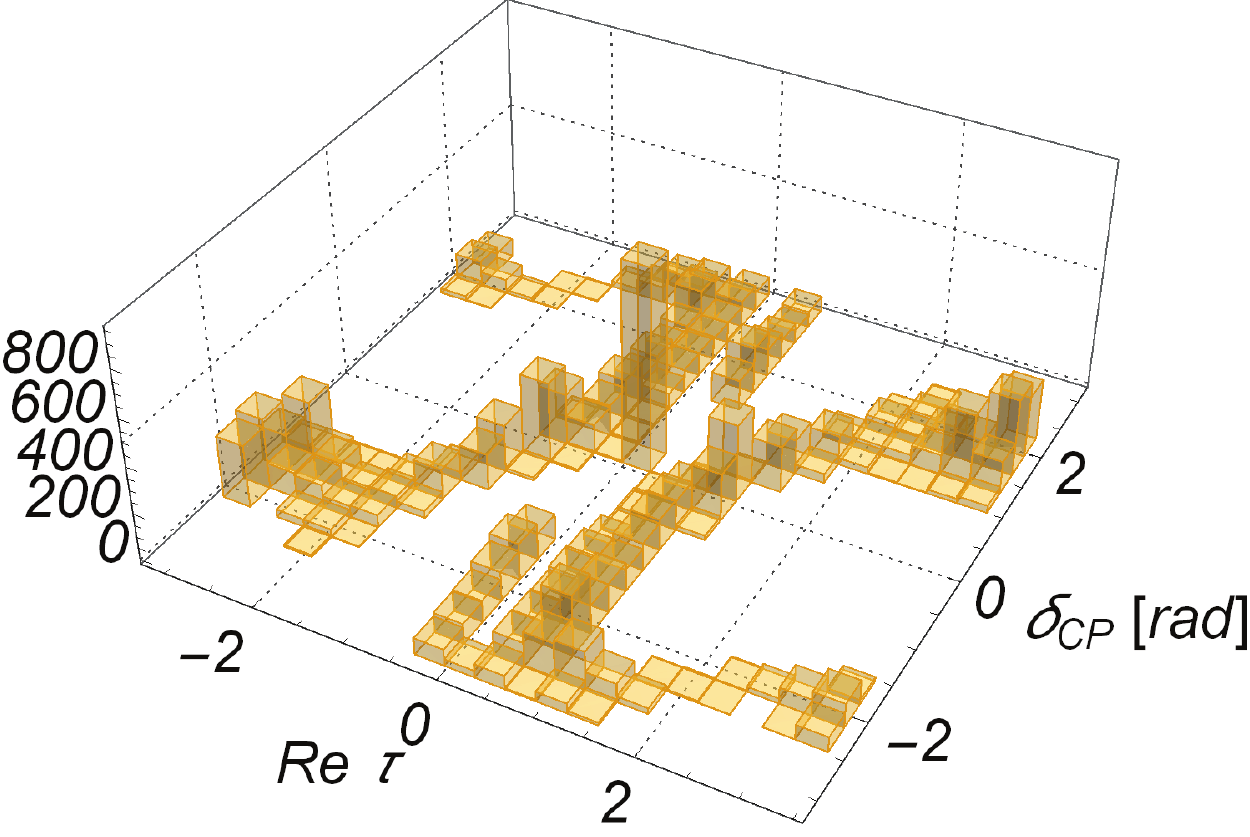}
\end{minipage}
\caption{Distributions of the CP-violating phase $\delta_{\rm CP}$ vs. real part of complex structure modulus ${\rm Re}\, \tau$ for ${\rm Im} \, \tau=2.0$ in Pattern I.}
\label{pattern1_2}
\end{figure}

{\bf Numerical analysis in Pattern I.}
Results are shown in {Figures}~\ref{pattern1_1} (for $\text{Im}\,\tau = 1.8$) and \ref{pattern1_2} (for $\text{Im}\,\tau = 2.0$), {which} indicate that it is not so trivial to realize $\delta_{\rm CP} \simeq 1.2 \, [{\rm rad}]$.
Note that in this {pattern}, the origin of the CP-violating phase is only a real part of the complex structure modulus parameter, i.e., ${\rm Re} \, \tau$ {since all the SS phases are set to be zero}.
{Figures}~\ref{pattern1_1} and \ref{pattern1_2} tell us that the quark flavor structure, i.e., the mass hierarchies and small mixing angles of the SM quarks, is {quite} dependent on the values of ${\rm Re}\, \tau$ as well as the values of ${\rm Im} \, \tau$.
Thus, we cannot fit the CP-violating phase by scattering ${\rm Re}\, \tau$ and the mass hierarchies and mixing angles by scattering ${\rm Im}\, \tau$, independently, and hence we need comprehensive {analyses} for the quark flavor structures by means of all of scattered Higgs VEVs, ${\rm Re}\, \tau$ and ${\rm Im}\, \tau$.
%On the other hand, we found that the parameter regions around $({\rm Re}\, \tau, {\rm Im}\, \tau) = (-3.0, 1.8), (-0.5, 1.8)$ [for $\text{Im}\,\tau = 1.8$] and $({\rm Re}\, \tau, {\rm Im}\, \tau) = (-0.5, 2.0), (2.0, 2.0), (3.0, 2.0)$ [for $\text{Im}\,\tau = 2.0$] lead to (quasi-)realizations of the observed value of $\delta_{\rm CP}$.

\begin{figure}[H]
\begin{minipage}{0.5\hsize}
\centering
\includegraphics[clip, width=0.8\hsize]{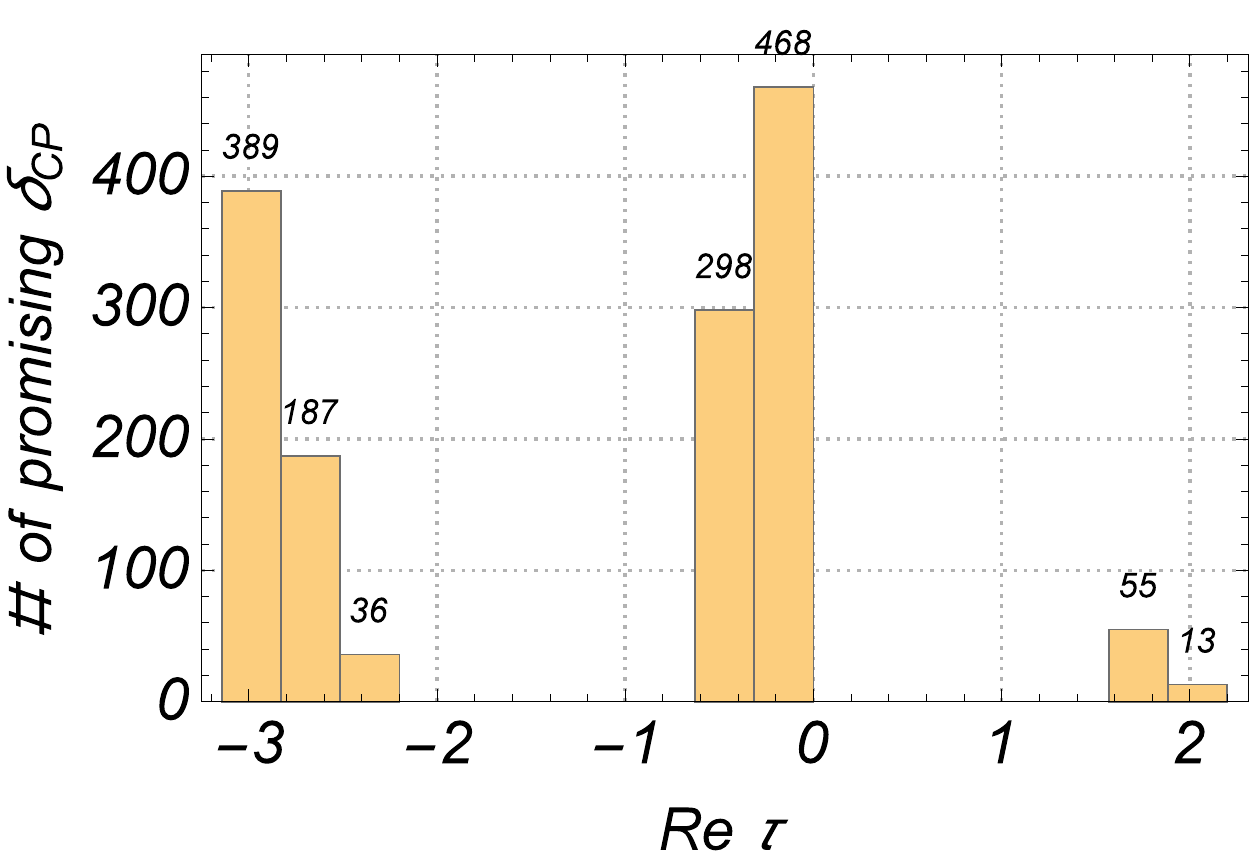}
\end{minipage}
\begin{minipage}{0.5\hsize}
\centering
\includegraphics[clip, width=0.8\hsize]{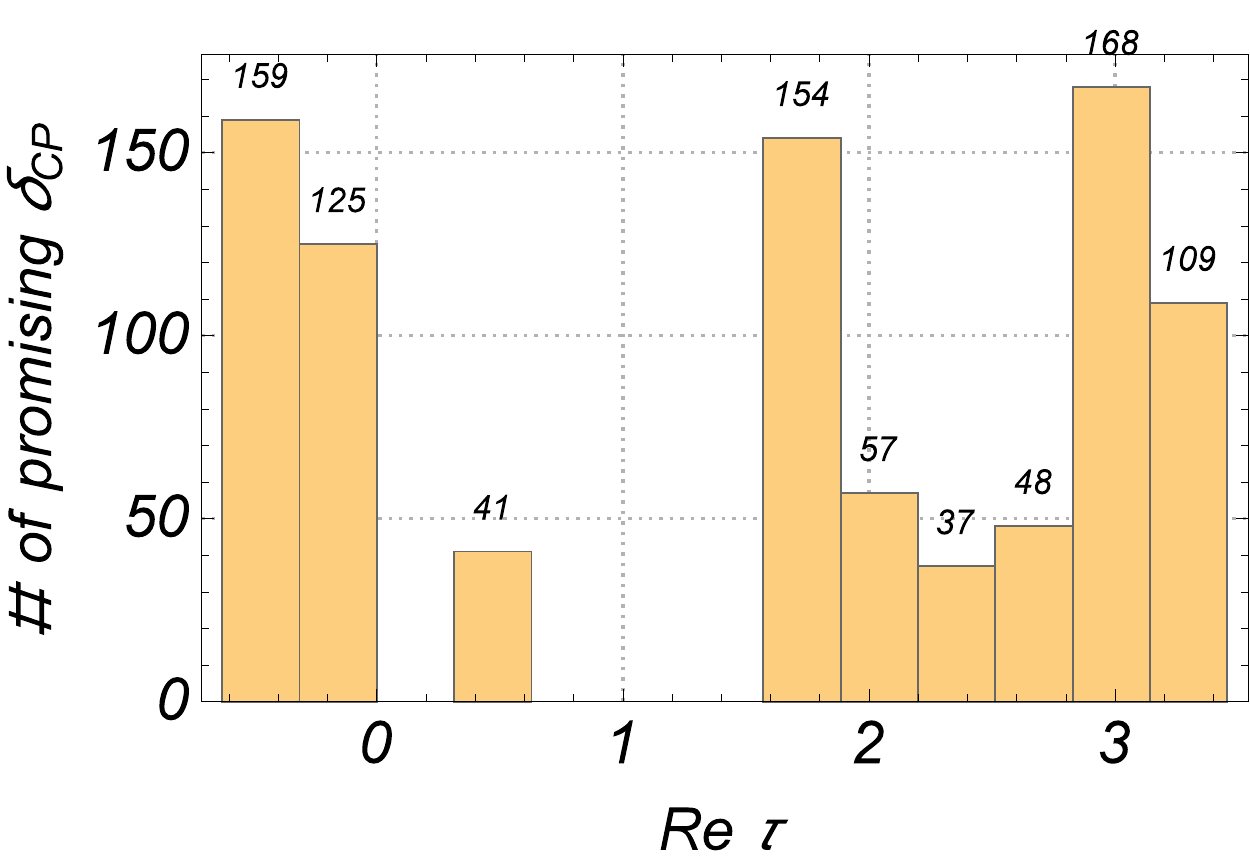}
\end{minipage}
\caption{Left panel: frequency of ${\rm Re}\, \tau$ satisfying an inequality in \eq{deltacpcut} for ${\rm Im} \, \tau=1.8$ in Pattern I.
{Digits on the top of} histogram bins denote the numbers of combinations of Higgs VEVs. Right panel: the same one for ${\rm Im} \, \tau=2.0$.}
\label{pattern1_3}
\end{figure}

We count the number of the configurations where values of the CP-violating phase are located in the quasi-realistic range,
\begin{gather}
0.8 \times 1.208 \leq \delta_{\rm CP} \leq 1.2 \times 1.208 \quad {\rm [rad]},
\label{deltacpcut}
\end{gather}
where we allow $20\%$ deviations from {the} observed value shown in~\eq{deltacp}.
The distributions are as shown in {Figure}~\ref{pattern1_3}, where various values of $\text{Re}\,\tau$ result in realizing the observed quark profiles, including the CP phase $\delta_{\text{CP}}$ in Pattern I.
In {particular}, we find characteristic peaks in the left and right panels of {Figure}~\ref{pattern1_3} around ${\rm Re} \, \tau \sim -3$ and {$-0.3$} [for ${\rm Im} \, \tau=1.8$] and ${\rm Re} \, \tau \sim -0.5, 1.5$ and $3$ [for ${\rm Im}\, \tau=2.0$], respectively.
This difference between the positions of characteristic peaks also implies that an imaginary part of the complex structure modulus parameter affects the value of the CP-violating phase as well as the mass hierarchies of the quarks.

In addition to that, Pattern I and Pattern $\text{\GkIV}$ are more promising than {the others} in the point of how many configurations derive suitable $\delta_{\text{CP}}$ defined in Eq.~(\ref{deltacpcut}), as we will see later.
This is easily found from the number of allowed configurations after the three cuts \eqs{upmasscut}--\eqref{mixcut} and the {region} for $\delta_{\text{CP}}$ in Eq.~(\ref{deltacpcut}).
The method and procedure to analyze the CP-violating phase are the same as those for the other patterns.


\begin{figure}[H]
\begin{minipage}{0.5\hsize}
\centering
\includegraphics[clip, width=0.85\hsize]{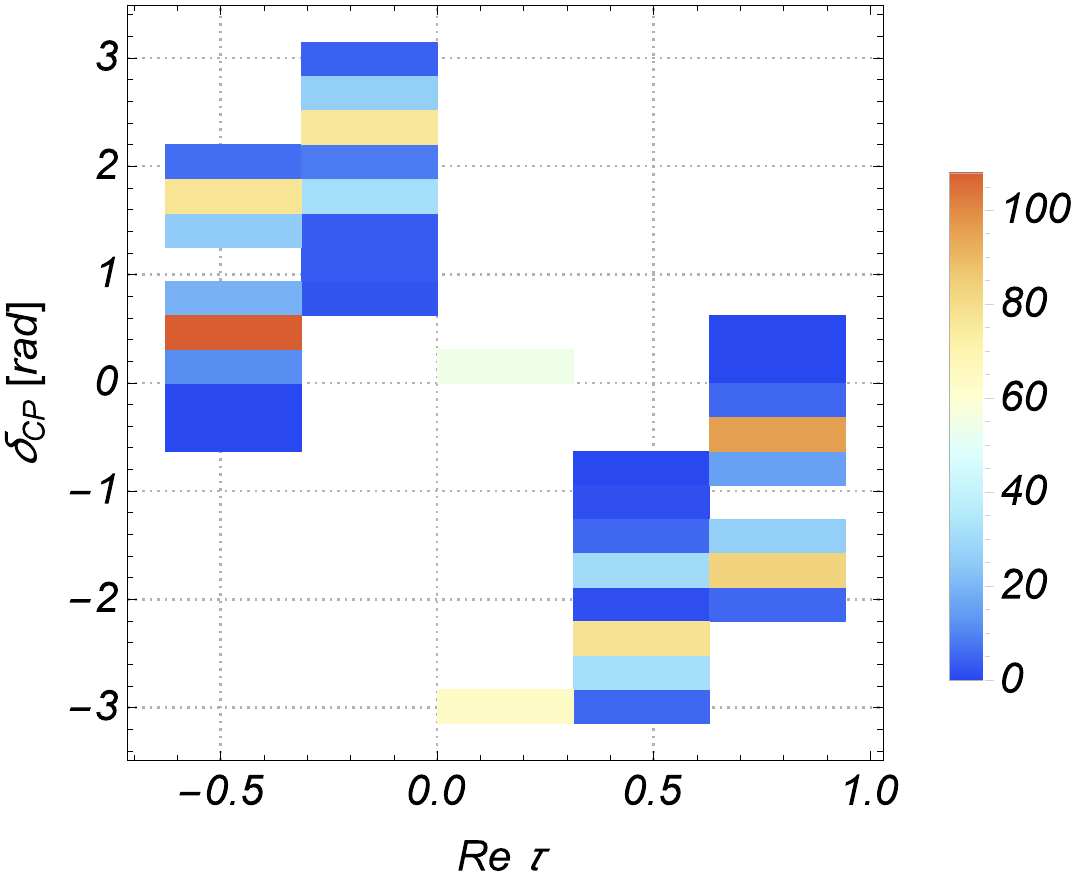}
\end{minipage}
\begin{minipage}{0.5\hsize}
\centering
\includegraphics[clip, width=0.85\hsize]{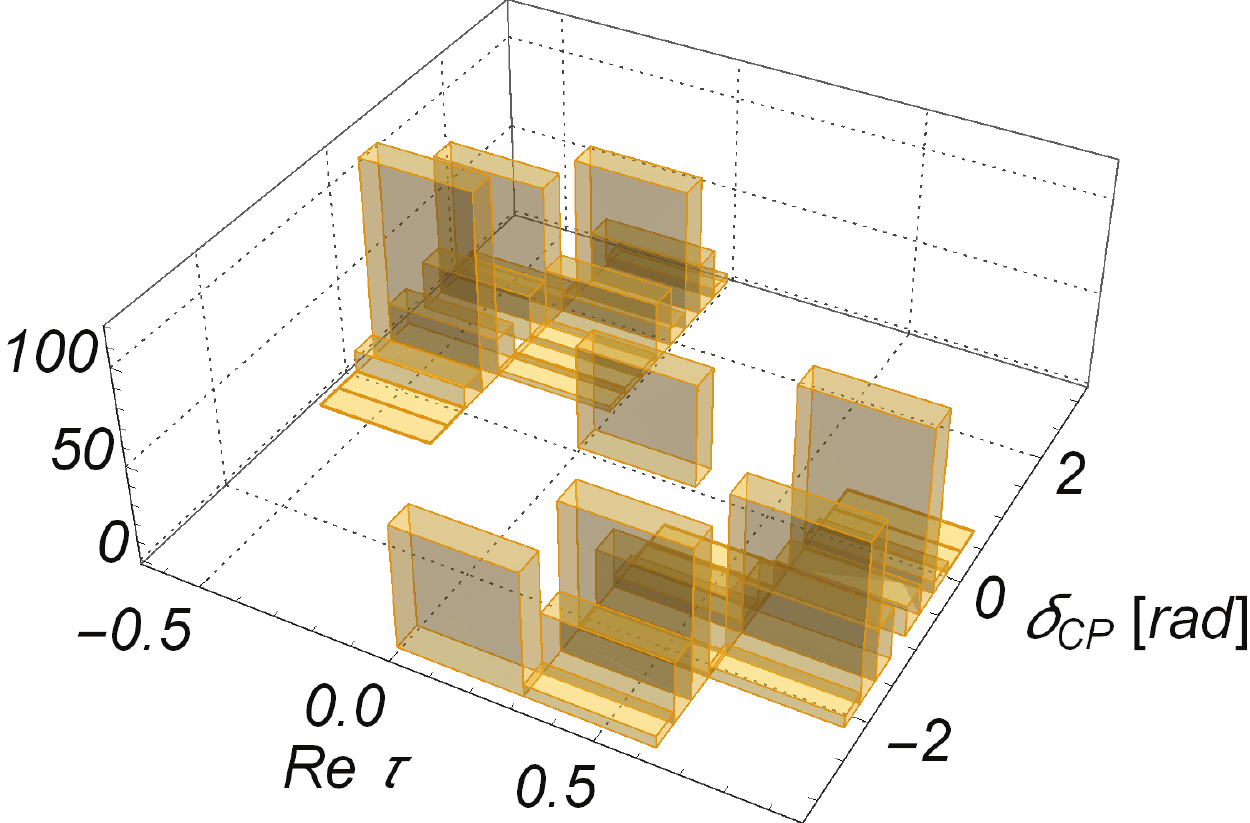}
\end{minipage}
\caption{Distributions of the CP-violating phase $\delta_{\rm CP}$ vs. real part of complex structure modulus ${\rm Re}\, \tau$ for ${\rm Im} \, \tau=1.7$ in Pattern {\GkII}.}
\label{pattern2_1}
\end{figure}

\begin{figure}[H]
\begin{minipage}{0.5\hsize}
\centering
\includegraphics[clip, width=0.85\hsize]{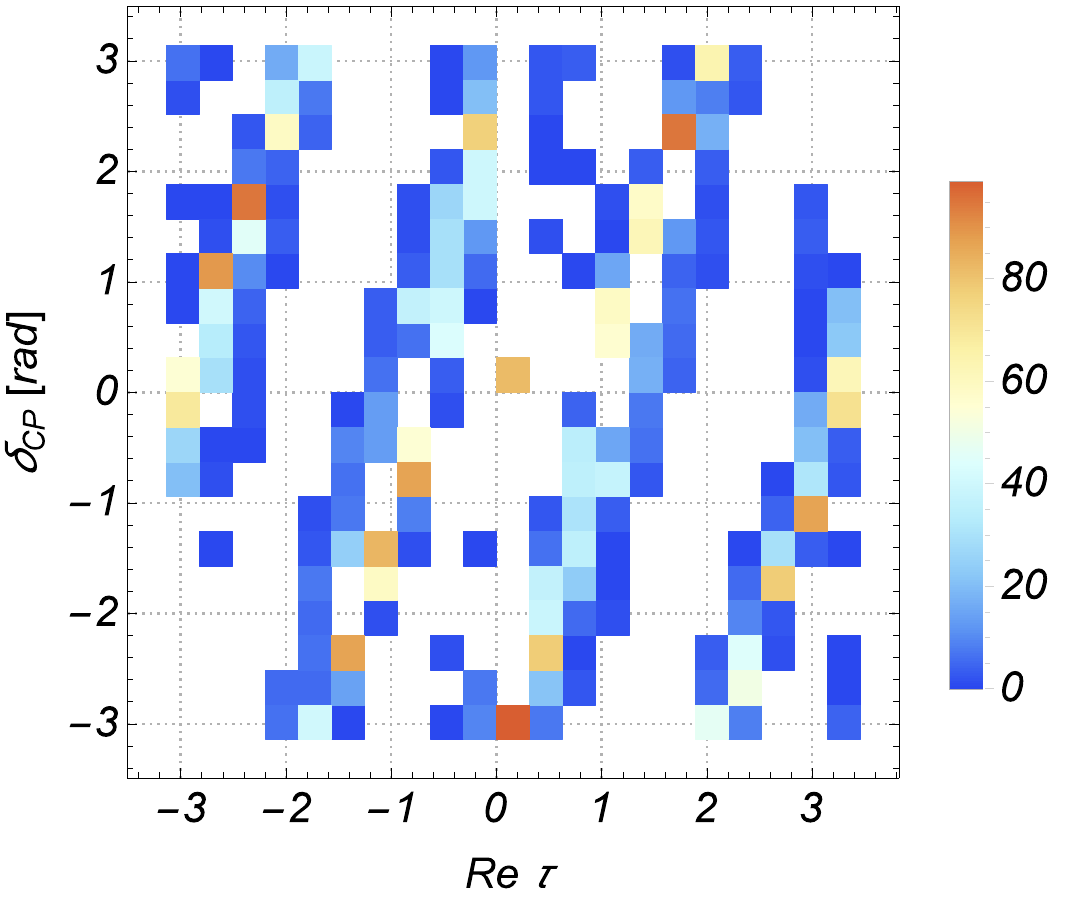}
\end{minipage}
\begin{minipage}{0.5\hsize}
\centering
\includegraphics[clip, width=0.85\hsize]{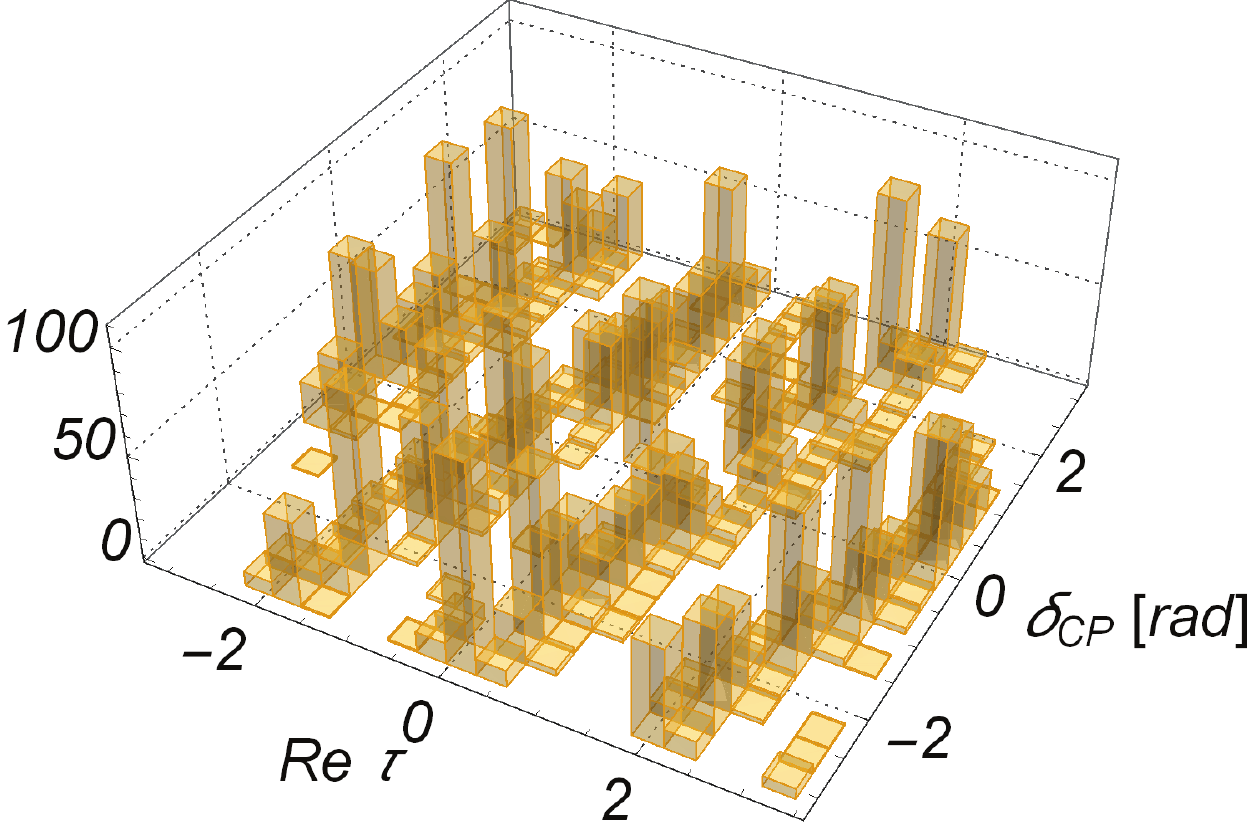}
\end{minipage}
\caption{Distributions of the CP-violating phase $\delta_{\rm CP}$ vs. real part of complex structure modulus ${\rm Re}\, \tau$ for ${\rm Im} \, \tau=1.9$ in Pattern {\GkII}.}
\label{pattern2_2}
\end{figure}

\begin{figure}[H]
\begin{minipage}{0.5\hsize}
\centering
\includegraphics[clip, width=0.8\hsize]{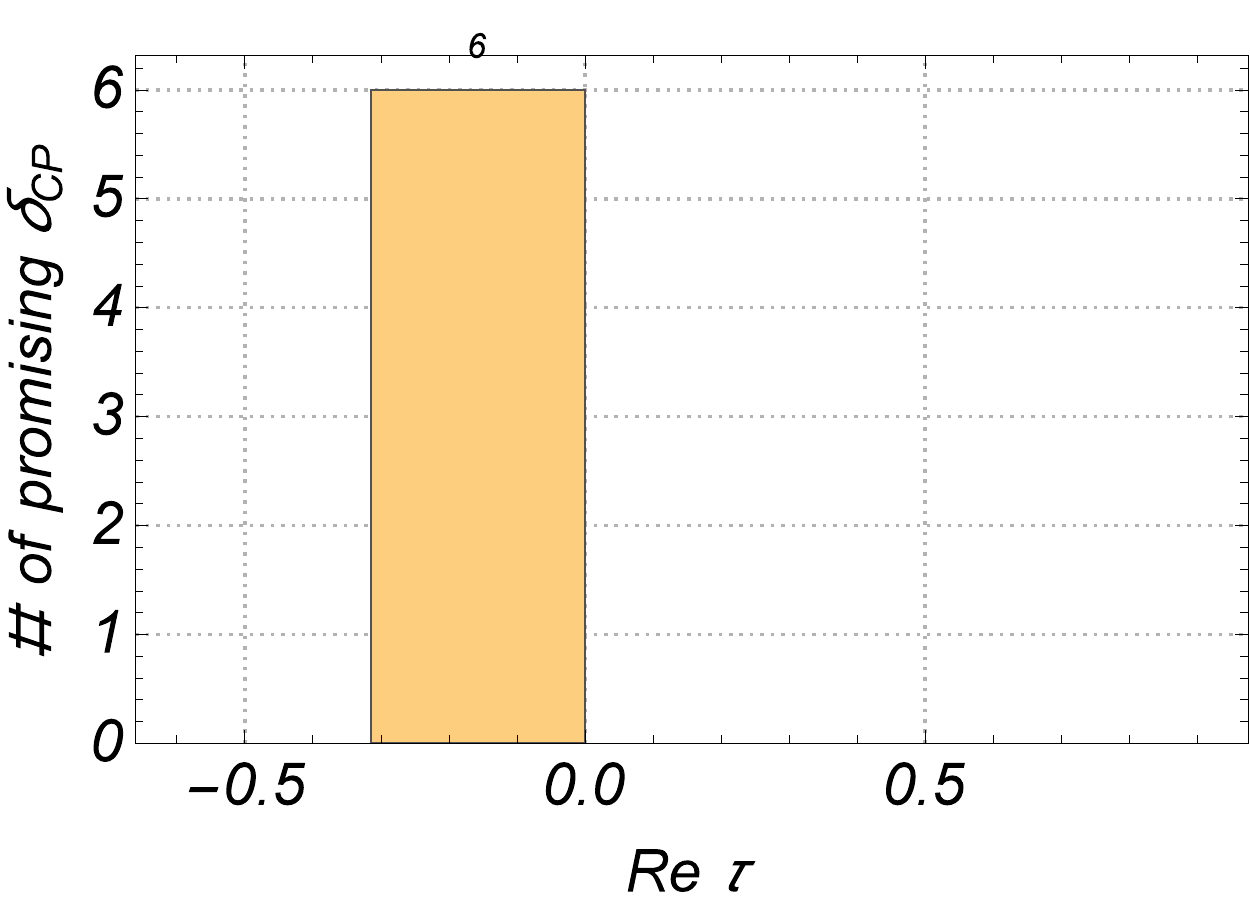}
\end{minipage}
\begin{minipage}{0.5\hsize}
\centering
\includegraphics[clip, width=0.8\hsize]{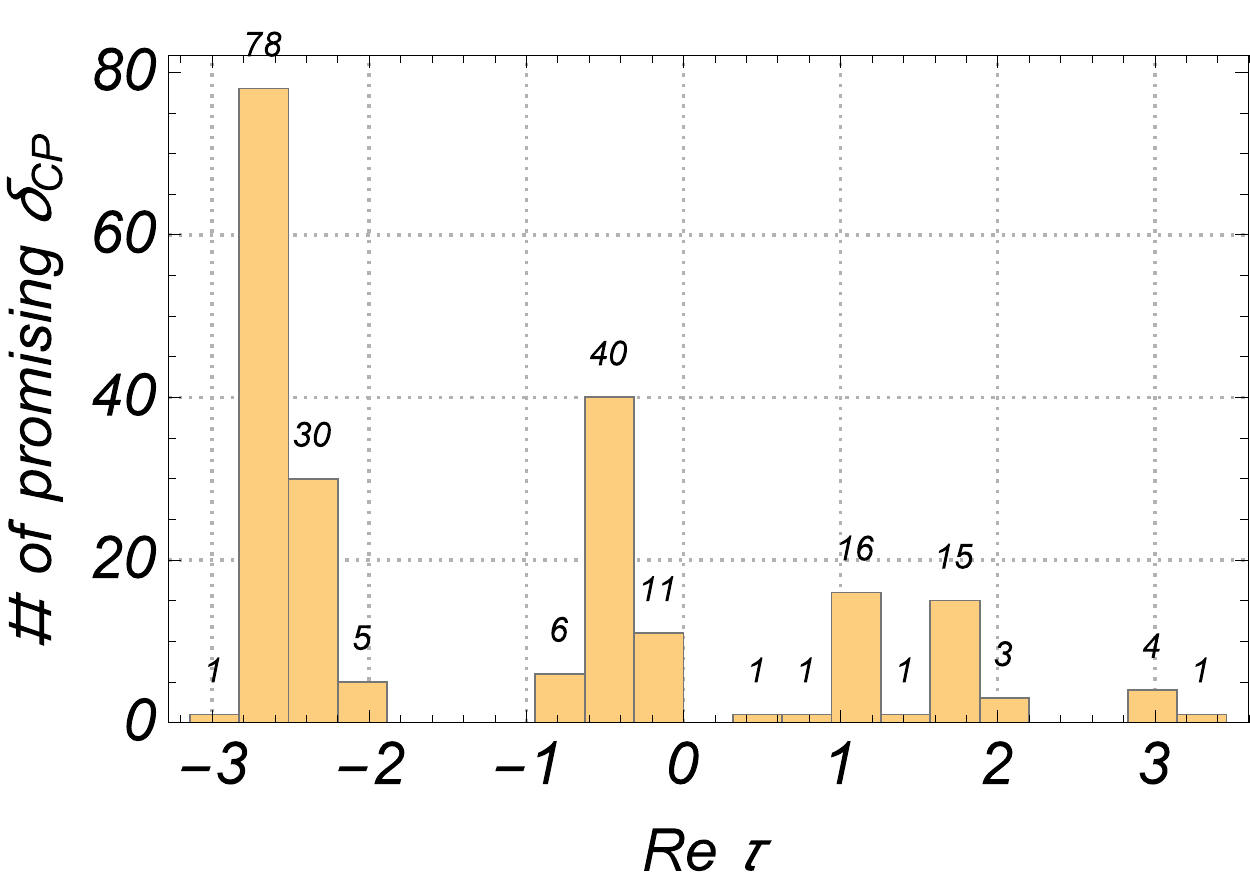}
\end{minipage}
\caption{Left panel: frequency of ${\rm Re}\, \tau$ satisfying an inequality in \eq{deltacpcut} for ${\rm Im} \, \tau=1.7$ in Pattern {\GkII}.
{Digits on the top of} histogram bins denote the numbers of combinations of Higgs VEVs. Right panel: the same one for ${\rm Im} \, \tau=1.9$.}
\label{pattern2_3}
\end{figure}

{\bf Numerical analysis in Pattern {\GkII}.}
Results are shown in {Figures}~\ref{pattern2_1} (for $\text{Im}\,\tau = 1.7$) and \ref{pattern2_2} (for $\text{Im}\,\tau = 1.9$) {and also \ref{pattern2_3} (histograms)}.
Note that in this {pattern}, the origin of the CP-violating phase is {not} only the complex-valued Yukawa couplings including the product of the non-vanishing ${\rm Re} \, \tau$, {but also} ``$a$'' part in the definition of the Jacobi's theta function \eqref{jacobi} {due to nonzero SS phases}.
{Figure}~{\ref{pattern2_3}} shows that to explain $\delta_{\rm CP} \simeq 1.2$ [rad]{, one} needs ${\rm Re}\, \tau \sim -0.3$ for ${\rm Im}\tau = 1.7$ and the result is more determinative.
On the other hand, {Figure}~\ref{pattern2_2} indicates that there exist various peaks in the right 3D histogram for ${\rm Im}\, \tau=1.9$, and also that we cannot extract a characteristic prediction in this case.
In addition, {an conclusion} revealed from {Figure}~\ref{pattern2_3} is that very {limited points are} only allowed by the numerical cuts.


\begin{figure}[H]
\begin{minipage}{0.5\hsize}
\centering
\includegraphics[clip, width=0.85\hsize]{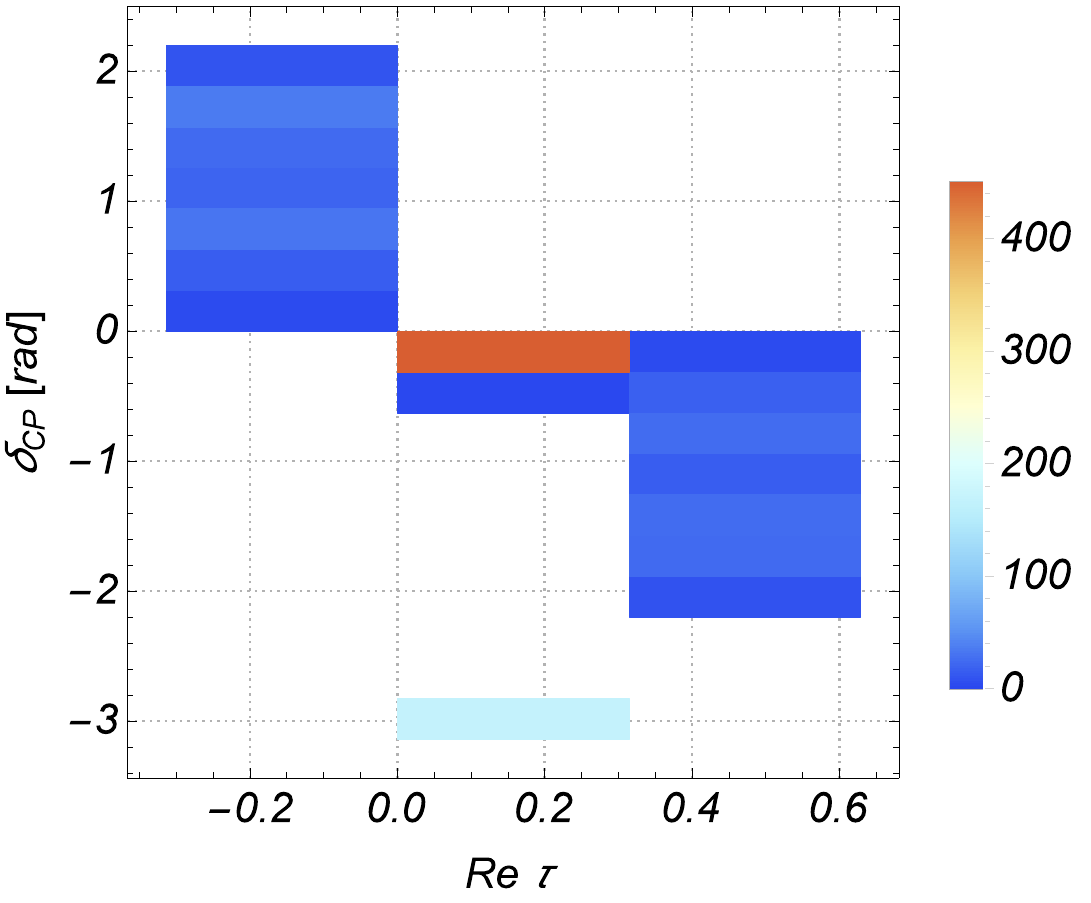}
\end{minipage}
\begin{minipage}{0.5\hsize}
\centering
\includegraphics[clip, width=0.85\hsize]{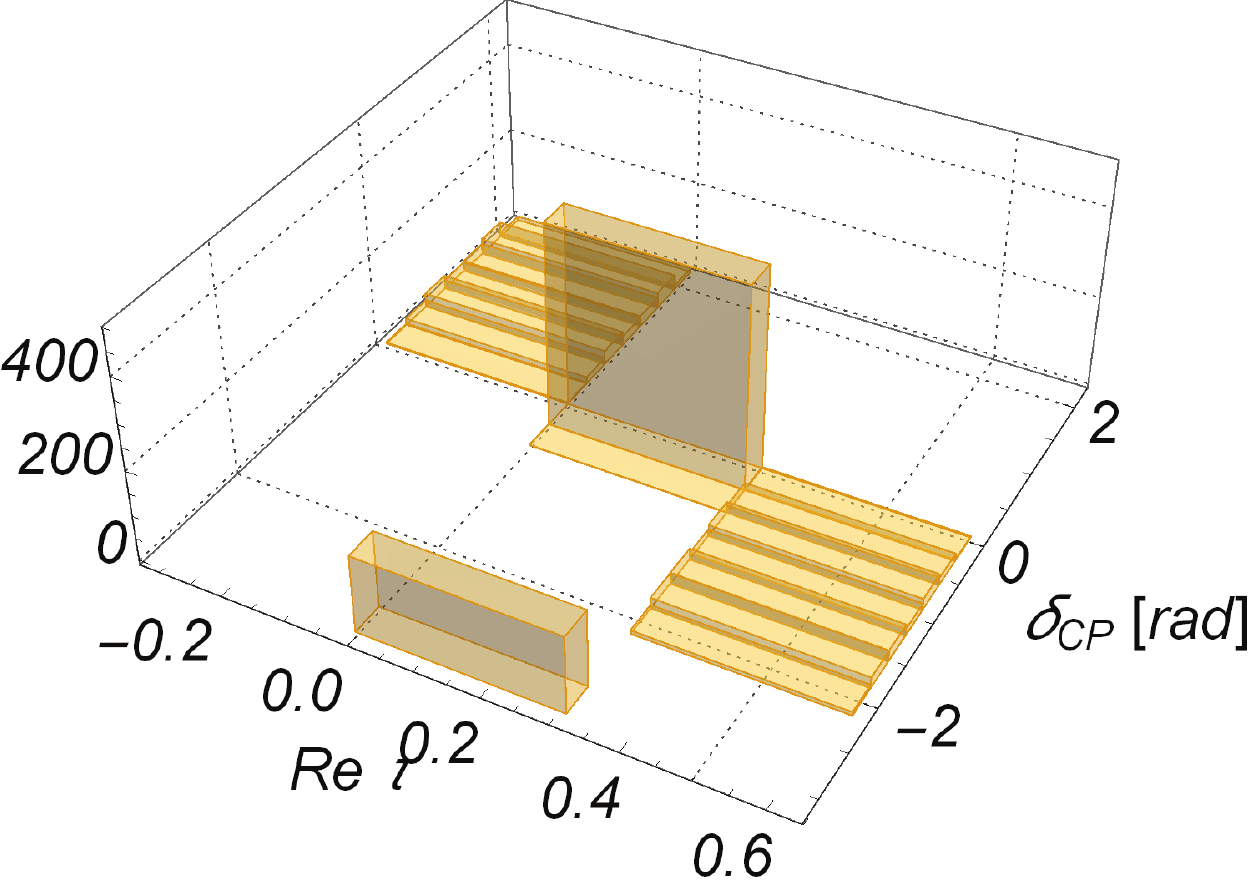}
\end{minipage}
\caption{Distributions of the CP-violating phase $\delta_{\rm CP}$ vs. real part of complex structure modulus ${\rm Re}\, \tau$ for ${\rm Im} \, \tau=1.9$ in Pattern {\GkIII}.}
\label{pattern3_1}
\end{figure}

\begin{figure}[H]
\begin{minipage}{0.5\hsize}
\centering
\includegraphics[clip, width=0.85\hsize]{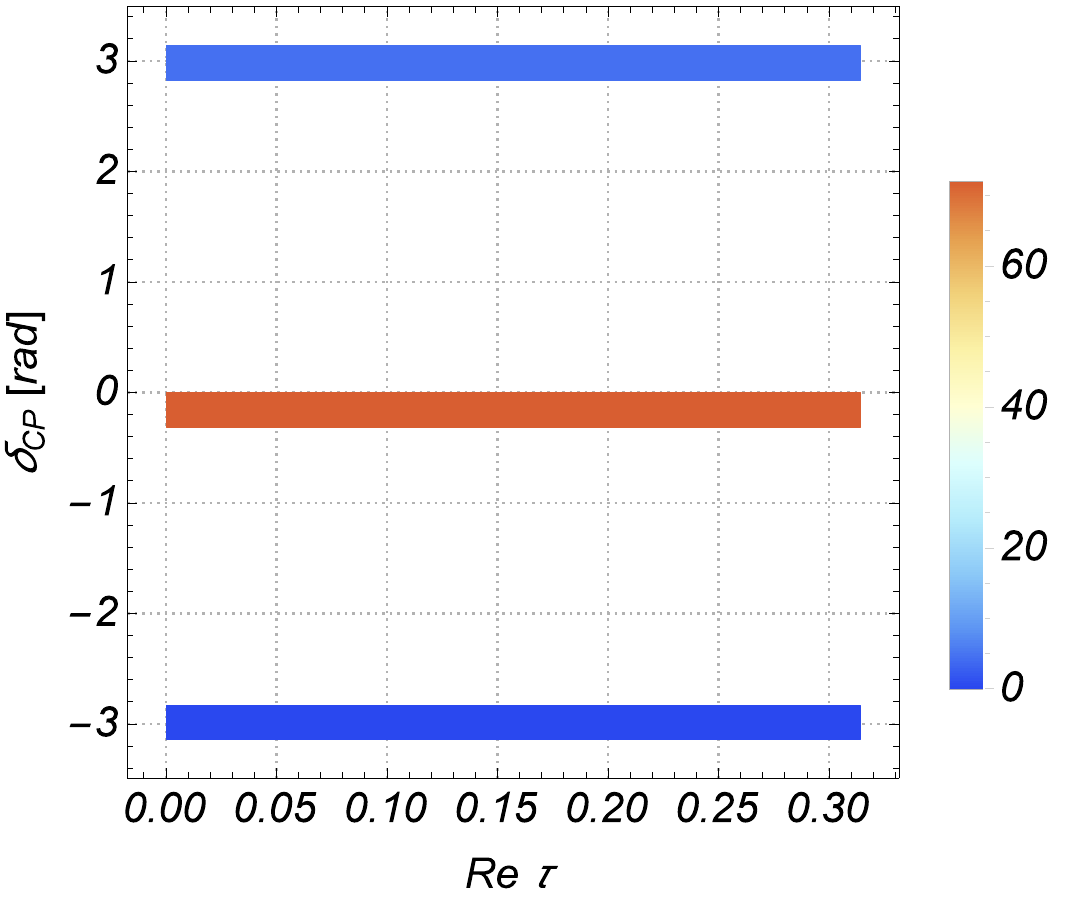}
\end{minipage}
\begin{minipage}{0.5\hsize}
\centering
\includegraphics[clip, width=0.85\hsize]{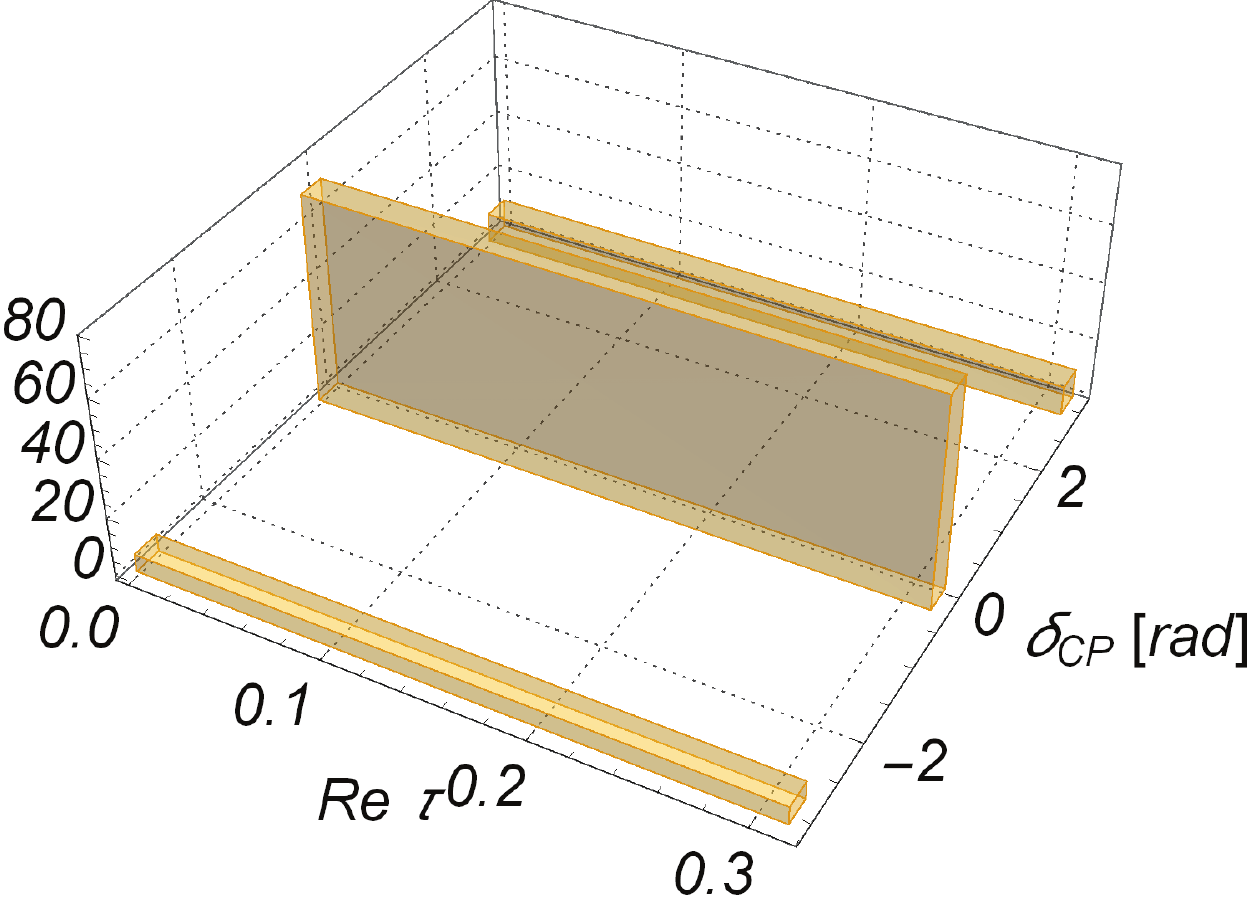}
\end{minipage}
\caption{Distributions of the CP-violating phase $\delta_{\rm CP}$ vs. real part of complex structure modulus ${\rm Re}\, \tau$ for ${\rm Im} \, \tau=2.0$ in Pattern {\GkIII}.}
\label{pattern3_2}
\end{figure}

\begin{figure}[H]
\begin{minipage}{0.5\hsize}
\centering
\includegraphics[clip, width=0.8\hsize]{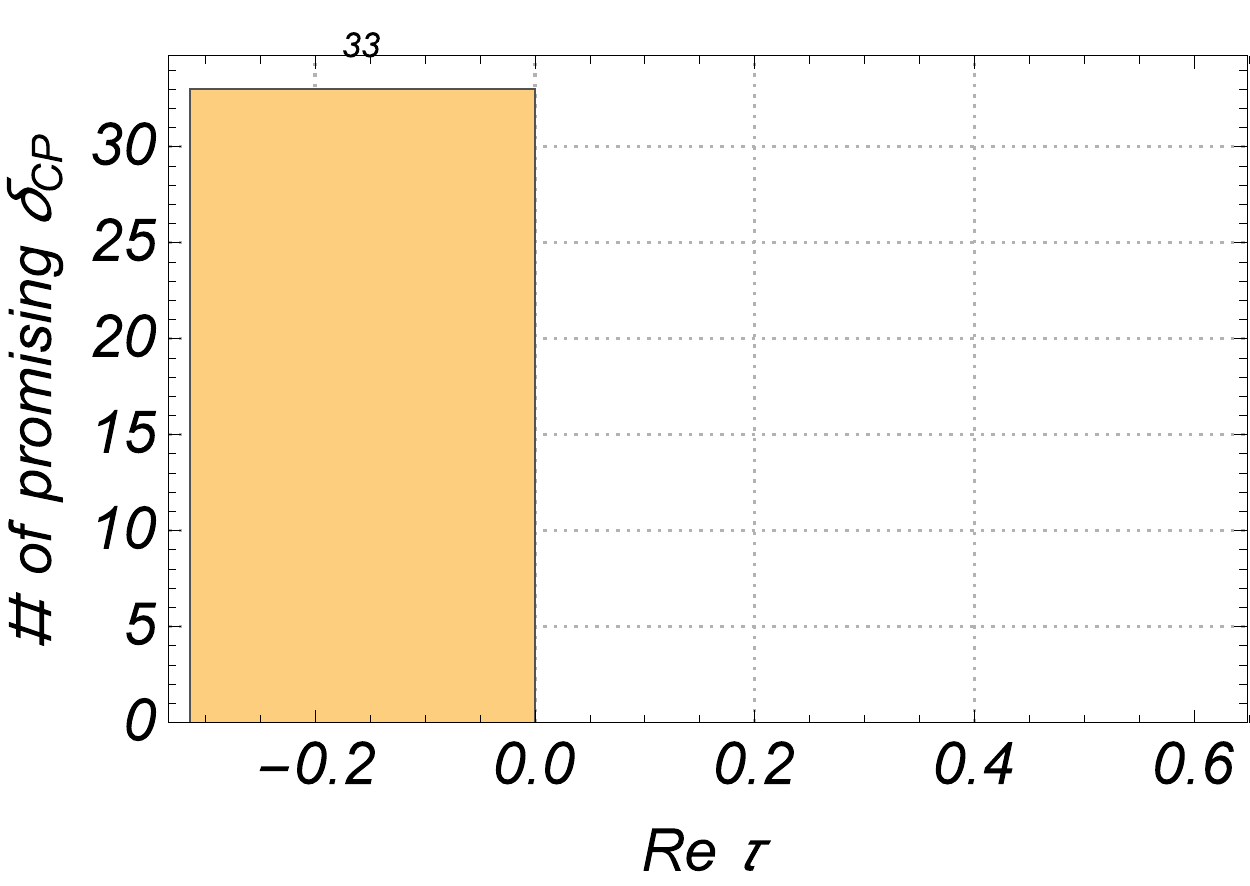}
\end{minipage}
\begin{minipage}{0.5\hsize}
\centering
\includegraphics[clip, width=0.8\hsize]{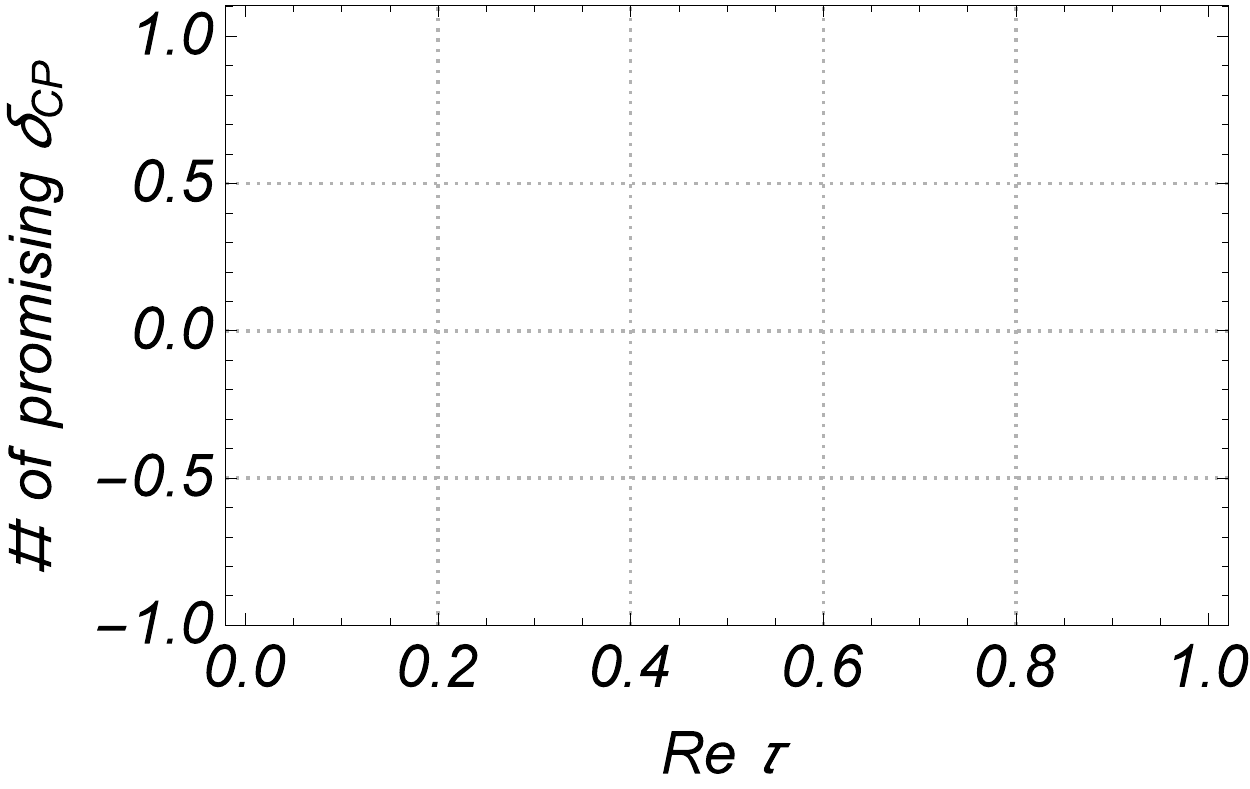}
\end{minipage}
\caption{Left panel: frequency of ${\rm Re}\, \tau$ satisfying an inequality in \eq{deltacpcut} for ${\rm Im} \, \tau=1.9$ in Pattern {\GkIII}.
{Digits on the top of} histogram bins denote the numbers of combinations of Higgs VEVs. Right panel: the same one for ${\rm Im} \, \tau=2.0$.}
\label{pattern3_3}
\end{figure}

{\bf Numerical analysis in Pattern {\GkIII}.}
Results are shown in {Figures}~\ref{pattern3_1} (for $\text{Im}\,\tau = 1.9$), \ref{pattern3_2} (for $\text{Im}\,\tau = 2.0$) and also \ref{pattern3_3} {(histograms)}.
Note that in this {pattern}, the origins of the CP-violating phase {is} the complex coefficient matrix \eqref{coefficients} {due to the SS phases,} as well as the complex-valued Yukawa couplings including the non-vanishing ${\rm Re} \, \tau$ and ``$c\nu$'' part in the definition of the Jacobi's theta function \eqref{jacobi}.
The case of ${\rm Im}\, \tau=2.0$ is drastic, because it is {impossible} to simultaneously explain the quark flavor structure of mass hierarchies and mixing angles in this case.


\begin{figure}[H]
\begin{minipage}{0.5\hsize}
\centering
\includegraphics[clip, width=0.85\hsize]{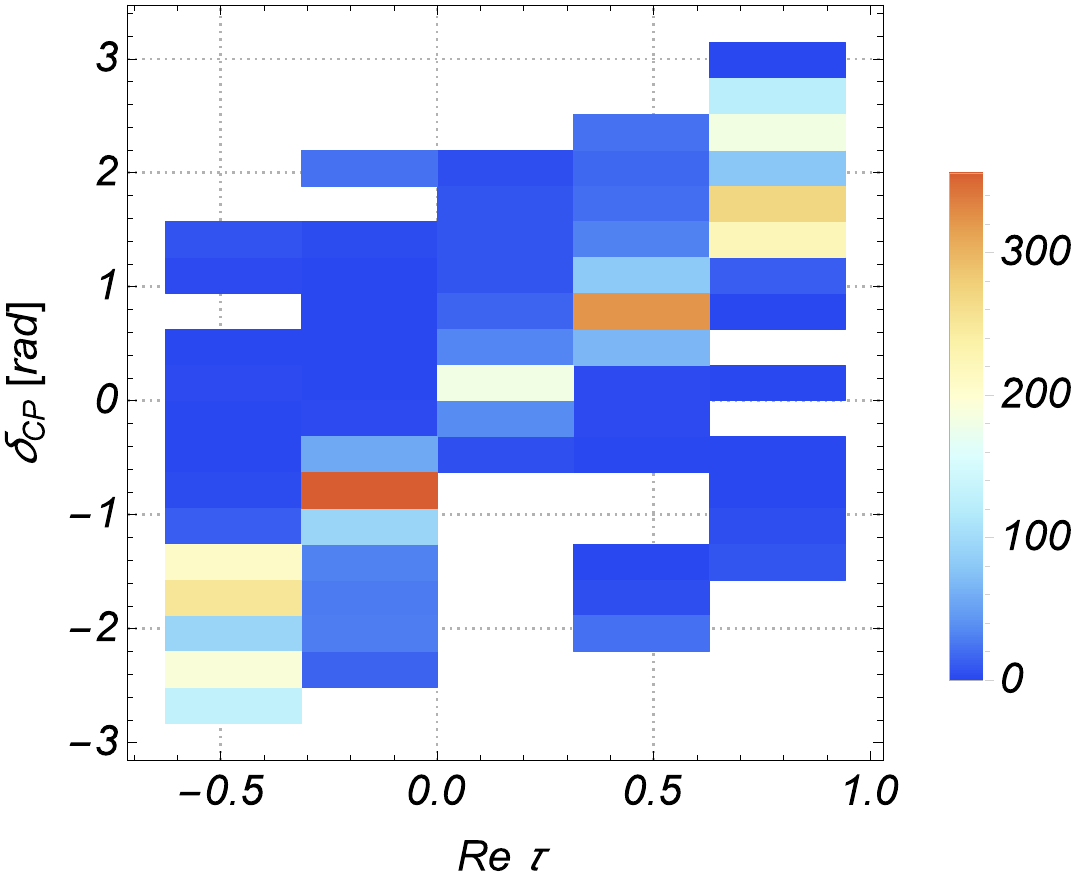}
\end{minipage}
\begin{minipage}{0.5\hsize}
\centering
\includegraphics[clip, width=0.85\hsize]{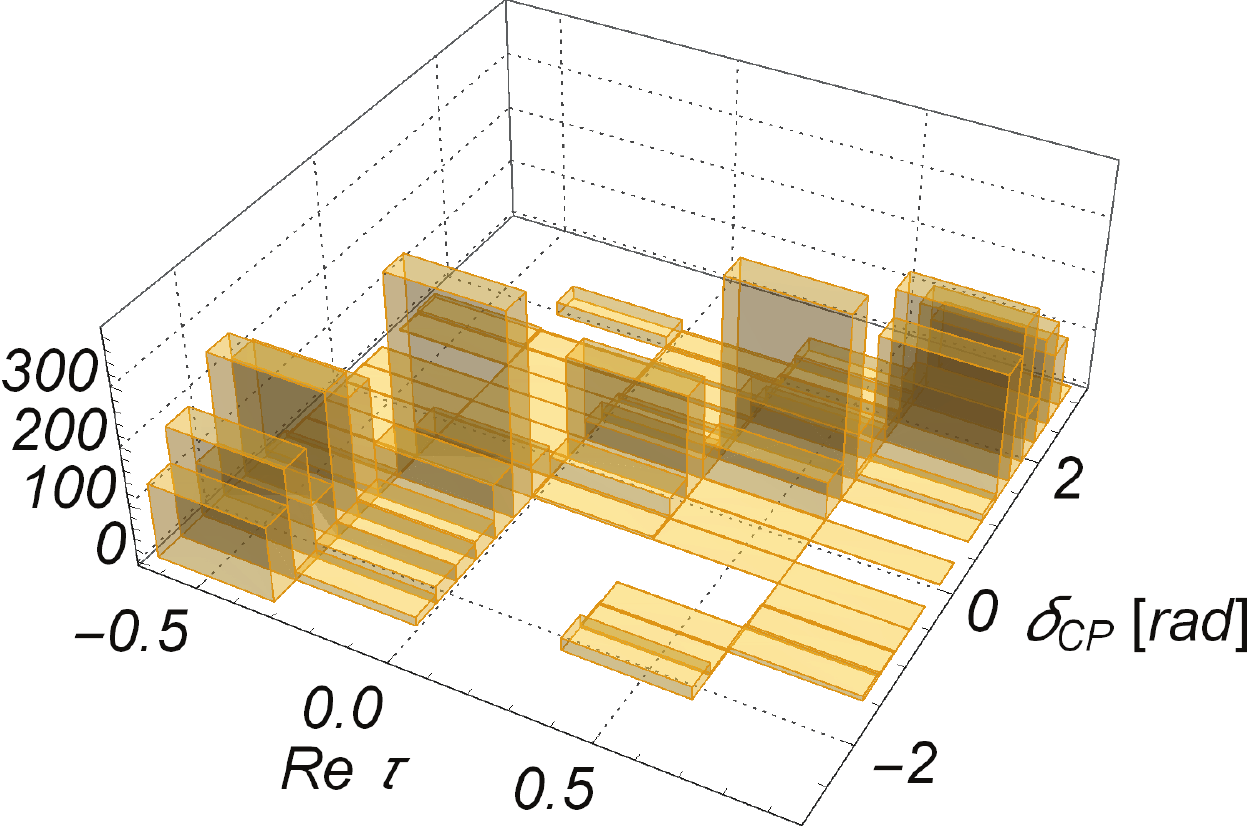}
\end{minipage}
\caption{Distributions of the CP-violating phase $\delta_{\rm CP}$ vs. real part of complex structure modulus ${\rm Re}\, \tau$ for ${\rm Im} \, \tau=1.7$ in Pattern {\GkIV}.}
\label{pattern4_1}
\end{figure}

\begin{figure}[H]
\begin{minipage}{0.5\hsize}
\centering
\includegraphics[clip, width=0.85\hsize]{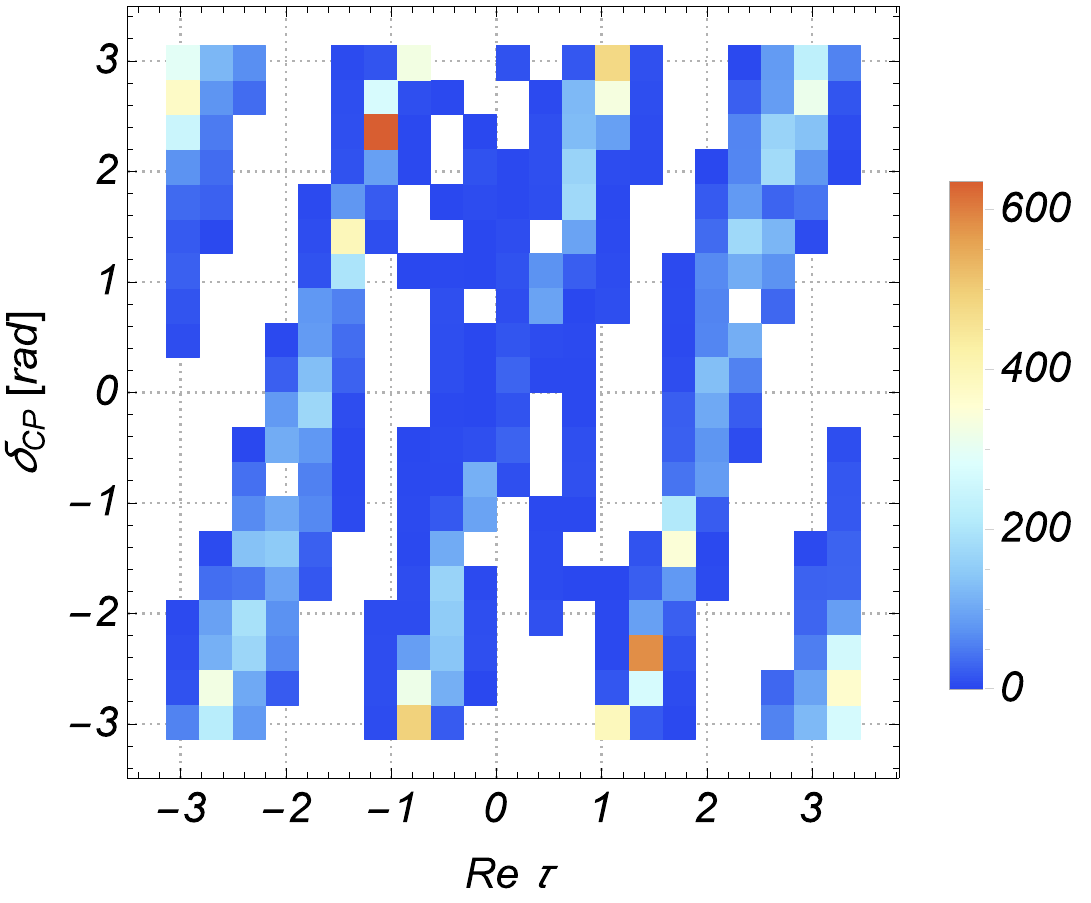}
\end{minipage}
\begin{minipage}{0.5\hsize}
\centering
\includegraphics[clip, width=0.85\hsize]{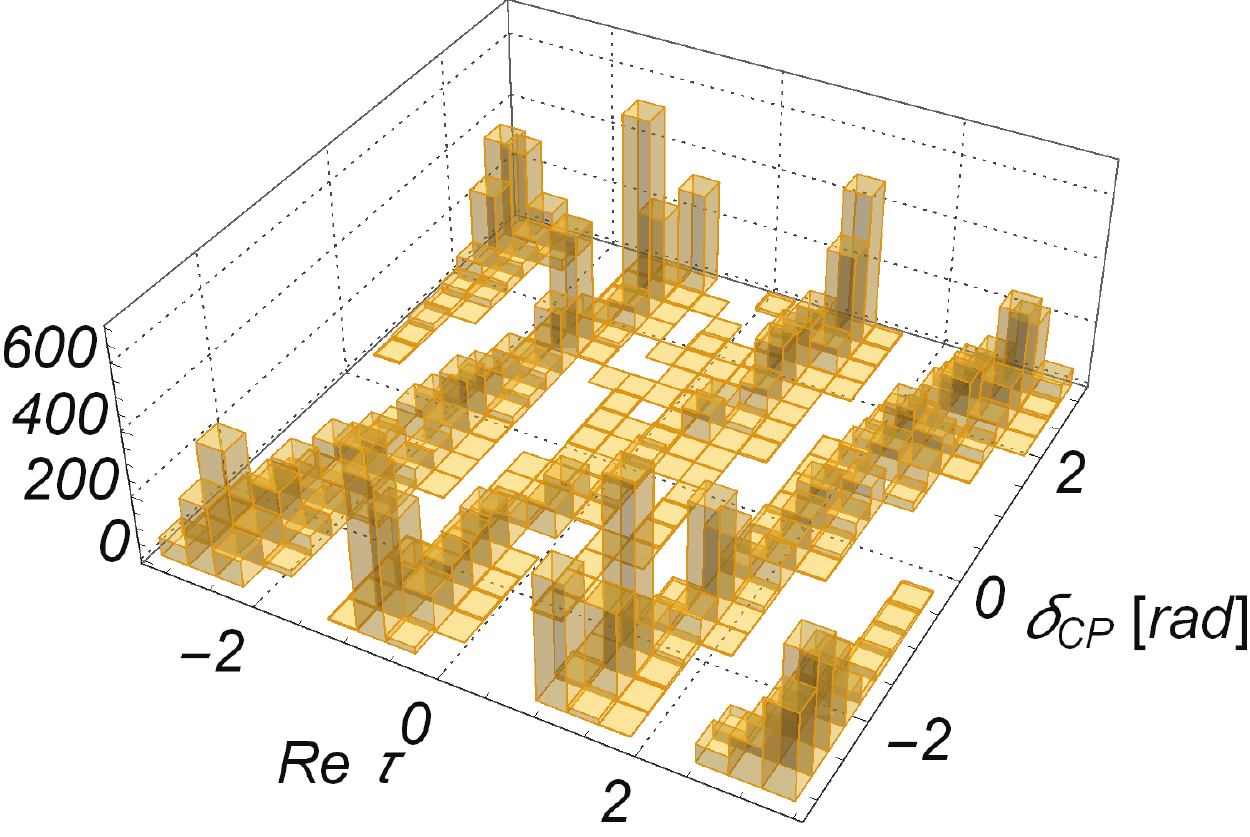}
\end{minipage}
\caption{Distributions of the CP-violating phase $\delta_{\rm CP}$ vs. real part of complex structure modulus ${\rm Re}\, \tau$ for ${\rm Im} \, \tau=1.9$ in Pattern {\GkIV}.}
\label{pattern4_2}
\end{figure}

\begin{figure}[H]
\begin{minipage}{0.5\hsize}
\centering
\includegraphics[clip, width=0.8\hsize]{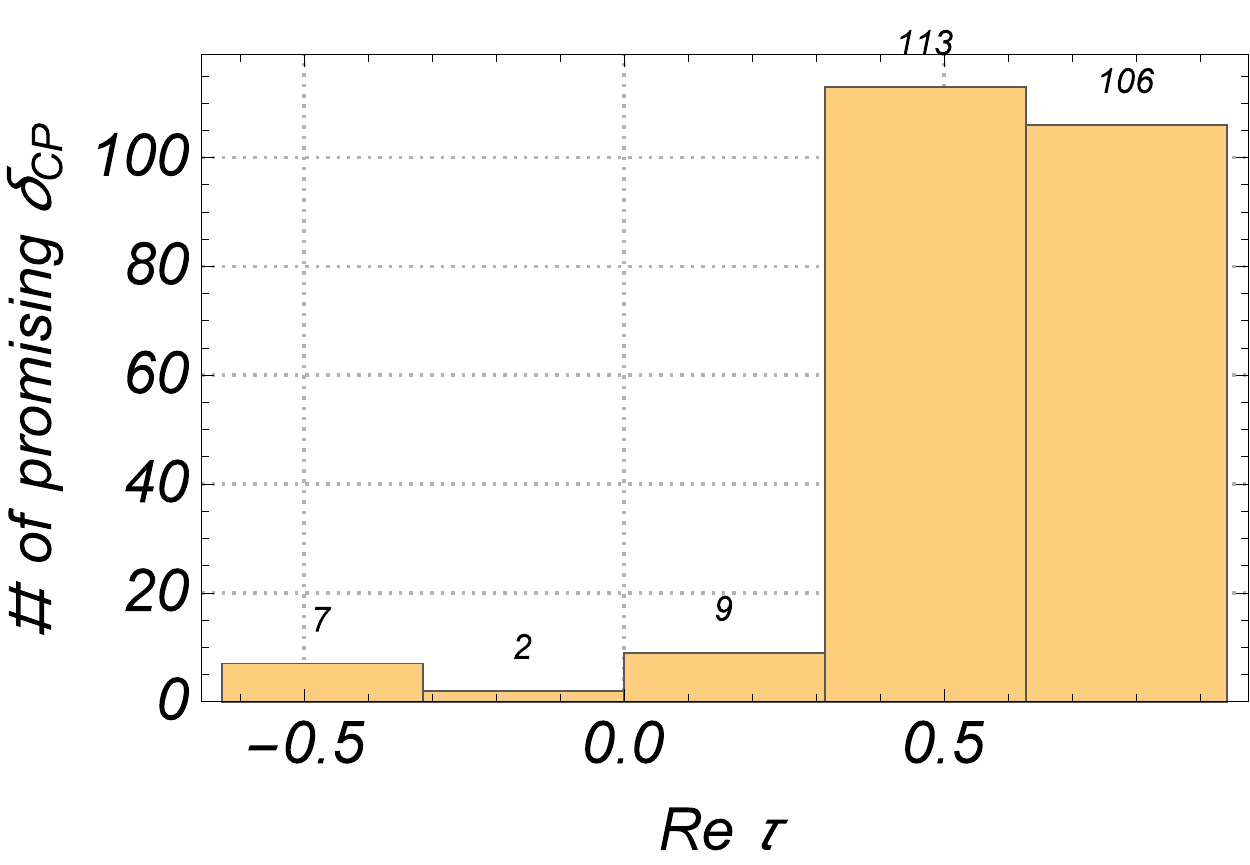}
\end{minipage}
\begin{minipage}{0.5\hsize}
\centering
\includegraphics[clip, width=0.8\hsize]{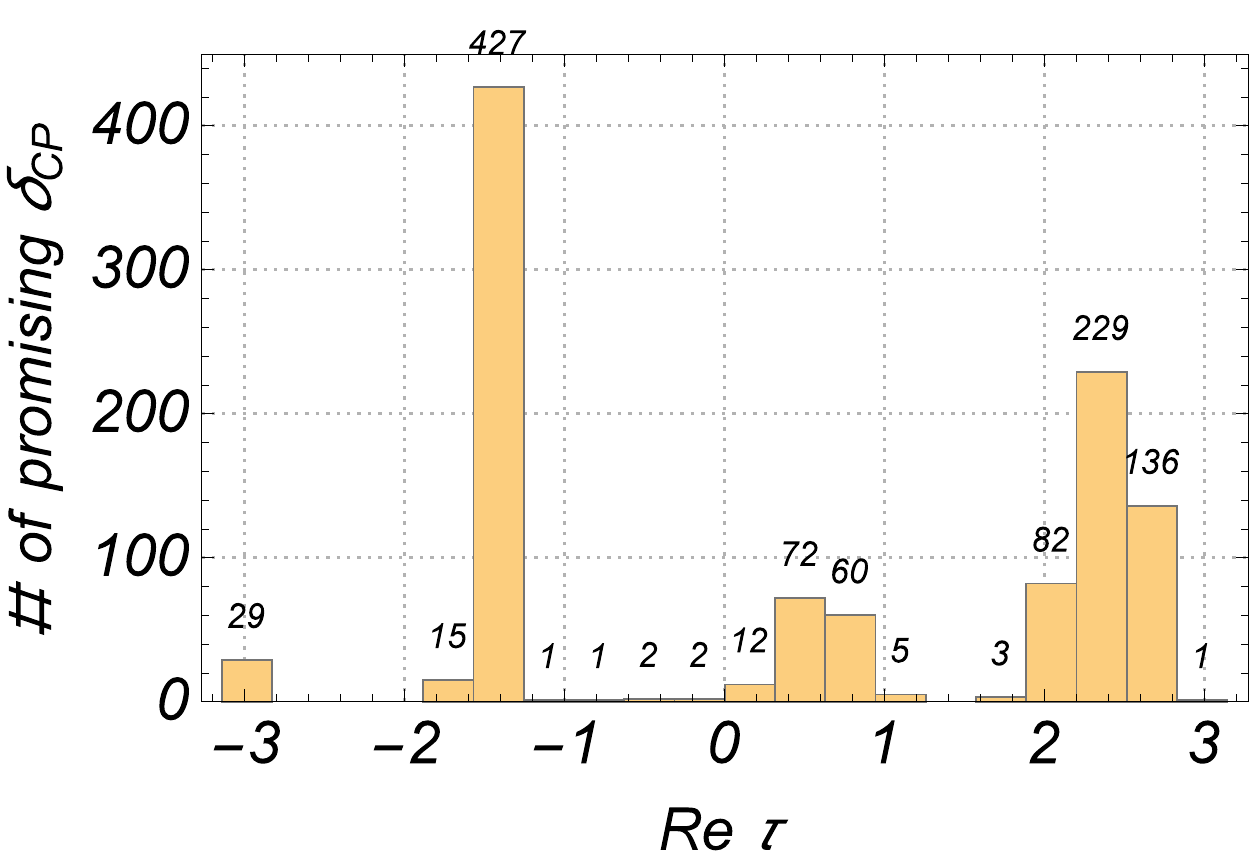}
\end{minipage}
\caption{Left panel: frequency of ${\rm Re}\, \tau$ satisfying an inequality in \eq{deltacpcut} for ${\rm Im} \, \tau=1.7$ in Pattern {\GkIV}.
{Digits on the top of} histogram bins denote the numbers of combinations of Higgs VEVs. Right panel: the same one for ${\rm Im} \, \tau=1.9$.}
\label{pattern4_3}
\end{figure}

{\bf Numerical analysis in Pattern {\GkIV}.}
Results are shown in {Figures}~\ref{pattern4_1} (for $\text{Im}\,\tau = 1.7$) and \ref{pattern4_2} (for $\text{Im}\,\tau = 1.9$) {and also \ref{pattern4_3} (histograms)}.
Note that in this {pattern}, {similar to the previous Pattern $\text{\GkIII}$,} the origin of the CP-violating phase comes from both of the complex-valued Yukawa couplings with the non-vanishing ${\rm Re}\, \tau$ and the coefficient matrices of two sectors ``1'' and ``2'', i.e., ${\cal U}^{Z_2: \eta_X} \, (X=1,2)$.
{Figure}~{\ref{pattern4_3}} indicates that ${\rm Re} \, \tau \sim$ {$0.5$--$0.75$} can lead to the observed value of the CP-violating phase, $\delta_{\rm CP} \simeq 1.2$ [rad], {where the detail of a sample point is shown in {Table} \ref{goodeg}}.
%, and also that the parameter range around ${\rm Re} \, \tau \sim 0.3--0.7$ is much more promising for ${\rm Im}\, \tau=1.7$, as the parameter range leads to a more realistic result of the quark flavor structure shown in {table} \ref{goodeg}.
{On the other hand}, the promising range of ${\rm Re}\, \tau$ for ${\rm Im} \, \tau=1.9$ is around ${\rm Re}\, \tau \sim -1.6$ and $2.2$.
This is easily understood also from the right panel in {Figure}~\ref{pattern4_2}.


%0
%%%%%%%%%%%%%%%%%%%%%%%%%%%%%%%%%%%%%%%%%%%%%%%%%%%
\section{Conclusions and discussions
\label{sec:summary}}
%%%%%%%%%%%%%%%%%%%%%%%%%%%%%%%%%%%%%%%%%%%%%%%%%%%

We have investigated properties of the CP-violating phase in the quark sector on toroidal orbifolds $T^2/Z_N \, (N=2,3,4,6)$ with non-vanishing magnetic fluxes.
In this system, a non-vanishing value is mandatory in the real part of the complex modulus parameter $\tau$ of the two torus in order to explain the CP violation in the quark sector.
On $T^2$ without orbifolding, underlying discrete flavor symmetries severely restrict the form of Yukawa couplings and it is very difficult to reproduce the observed pattern in the quark sector including the CP-violating phase $\delta_{\rm CP}$.
Only on the case of $T^2/Z_2$ orbifold with multiple Higgs doublets were focused, since under the other orbifoldings, i.e., $T^2/Z_N \, (N=3,4,6)$, the FN factor $e^{-{\rm Im}\, \tau}$ is not so small enough for causing the Gaussian FN mechanism.
We numerically analyzed four patterns of the SS phases and
as a result, it has been revealed that we can obtain realistic values near $\delta_{\rm CP} \sim 1.2 \, {[\rm rad],}$ by appropriate arrangement of the value of ${\rm Re}\, \tau$ for all of the patterns.

As concretely addressed in Ref.~\cite{Abe:2015yva,Fujimoto:2016zjs}, in the present framework of the magnetized $T^2/Z_2$ orbifolds, multiple Higgs doublets are mandatory for realizing the observed quark masses and mixing angles.
Here, let us comment on a possible problem via the existence of multiple Higgs doublets.
In general, if no discrete symmetry or accidental cancellation realized by tuning of parameters happens, sizable tree-level flavor-changing neutral interactions are induced (in mass eigenbasis), which are highly disfavored by experimental results.
To handle the danger without addressing symmetry or cancellation, except for the observed $125\,\text{GeV}$ Higgs boson, doublet scalars are requested to be massive as much as around $10^2 \sim 10^3\,\text{TeV}$.
Dynamical realizations of Higgs $\mu$ terms with one light direction via D-brane instanton effects would give us an actual prescription (see e.g.,~\cite{Abe:2015uma,Kobayashi:2015siy}).

%0
%\section*{Acknowledgment}
\begin{acknowledgments}
T.K. and Y.T. are supported in part by Grants-in-Aid for Scientific Research No.~26247042~{(T.K.)} and 
No.~16J04612~{(Y.T.)} from the Ministry of Education, Culture, Sports, Science and Technology~{(MEXT)} in Japan.
\end{acknowledgments}

%0
%\section*{{Appendix}}
\appendix
\section{The other configuration of magnetic fluxes
\label{appendix:others}}
\begin{figure}[H]
\begin{minipage}{0.5\hsize}
\centering
\includegraphics[clip, width=0.85\hsize]{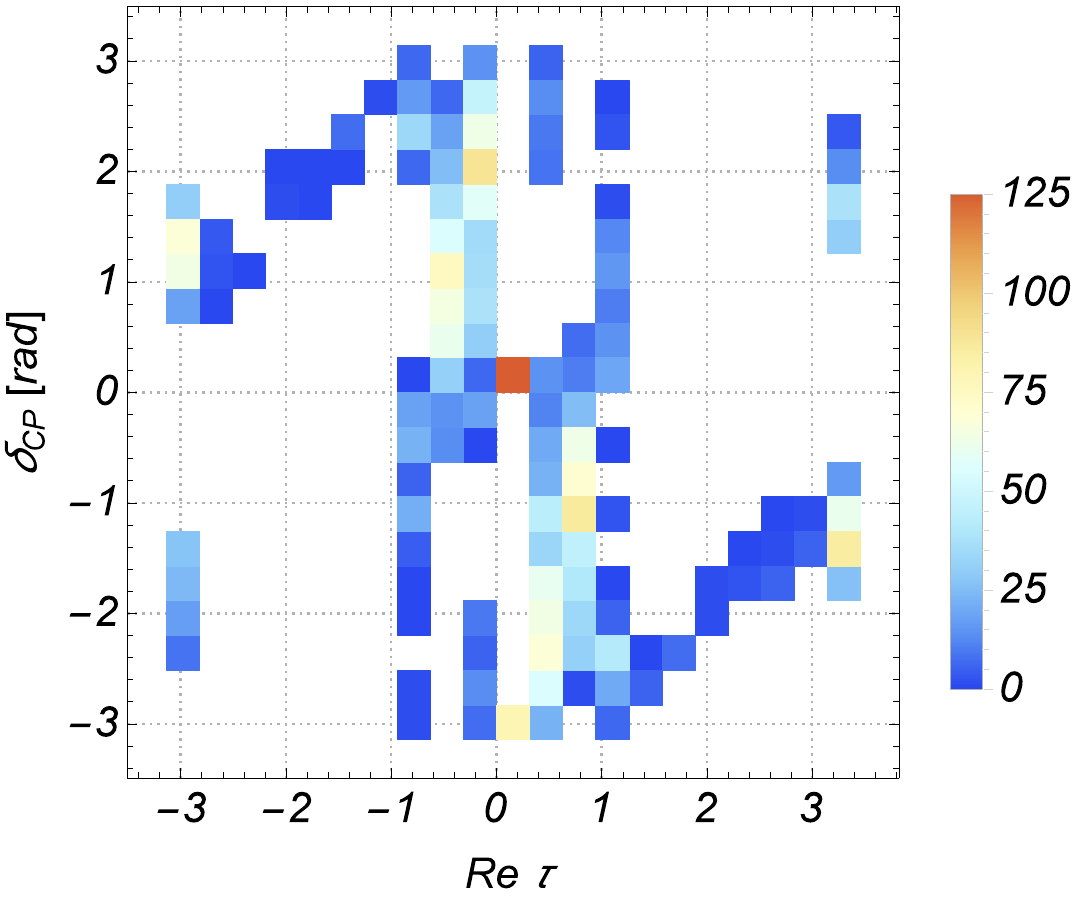}
\end{minipage}
\begin{minipage}{0.5\hsize}
\centering
\includegraphics[clip, width=0.85\hsize]{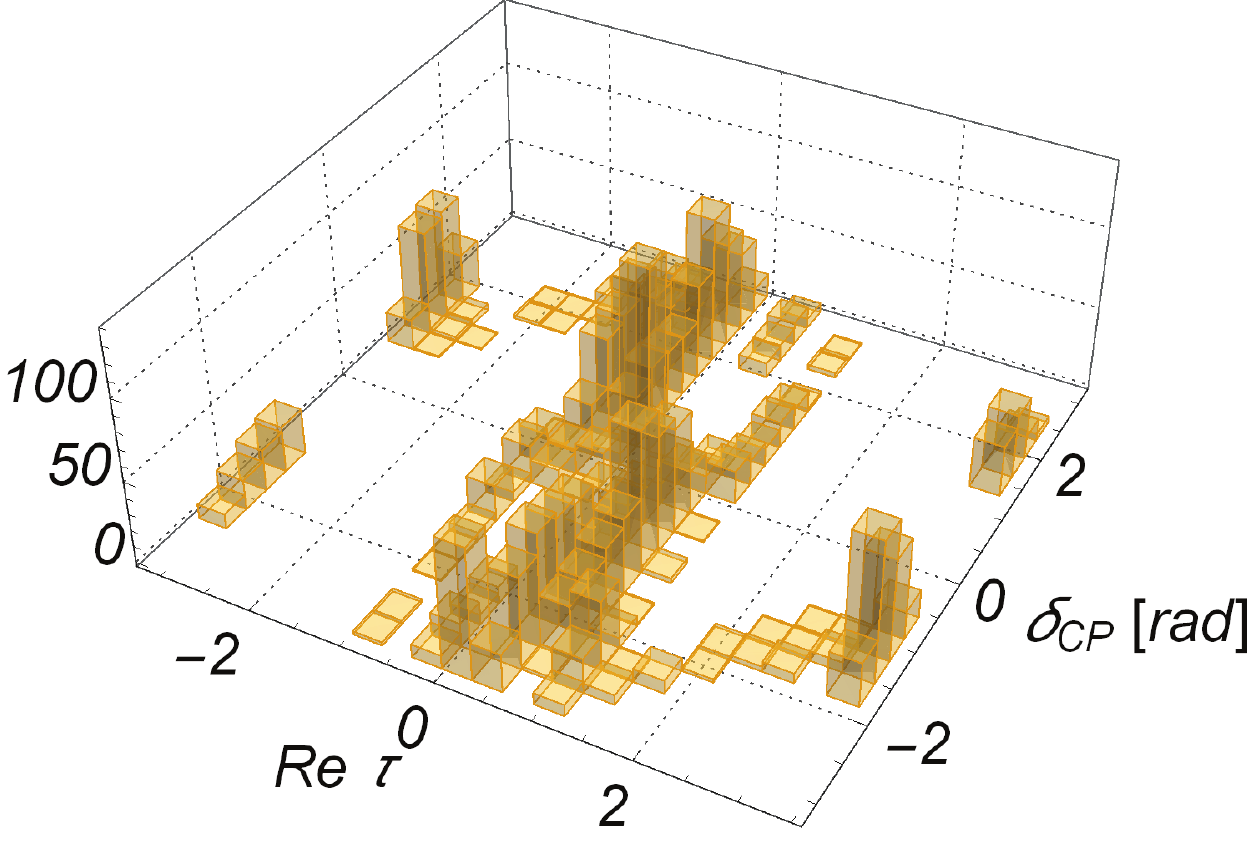}
\end{minipage}
\caption{Distributions of CP-violating phase $\delta_{\rm CP}$ vs. real part of complex structure modulus ${\rm Re}\, \tau$ for ${\rm Im} \, \tau=1.8$ in Pattern V.}
\label{pattern5_1}
\end{figure}

\begin{figure}[H]
\begin{minipage}{0.5\hsize}
\centering
\includegraphics[clip, width=0.85\hsize]{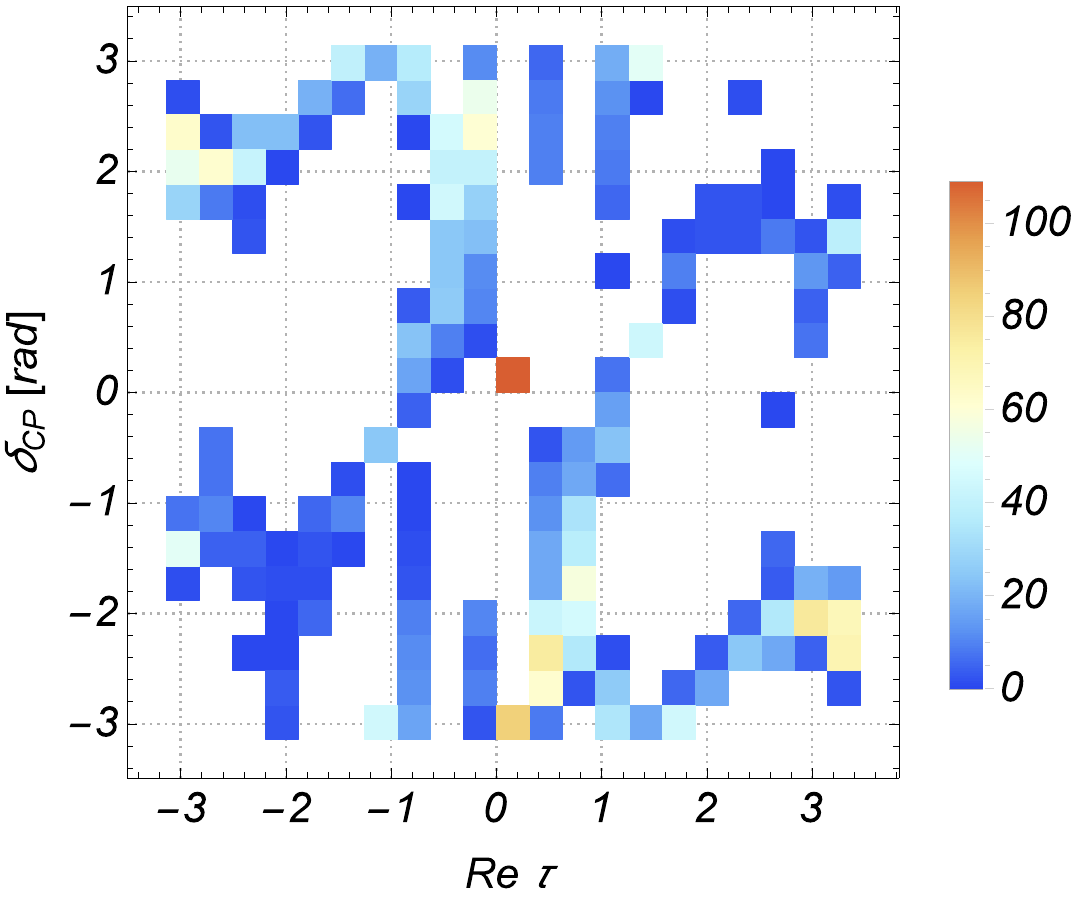}
\end{minipage}
\begin{minipage}{0.5\hsize}
\centering
\includegraphics[clip, width=0.85\hsize]{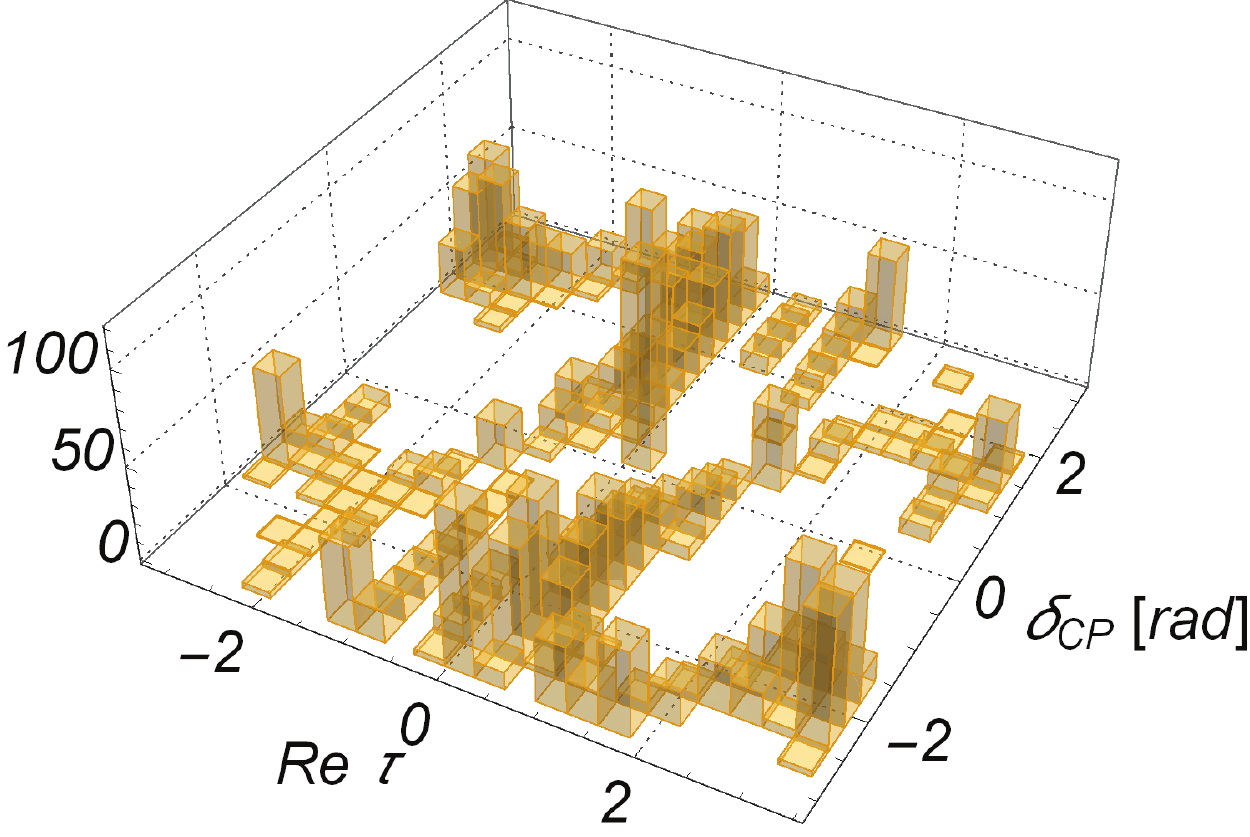}
\end{minipage}
\caption{Distributions of CP-violating phase $\delta_{\rm CP}$ vs. real part of complex structure modulus ${\rm Re}\, \tau$ for ${\rm Im} \, \tau=2.0$ in Pattern V.}
\label{pattern5_2}
\end{figure}

\begin{figure}[H]
\begin{minipage}{0.5\hsize}
\centering
\includegraphics[clip, width=0.85\hsize]{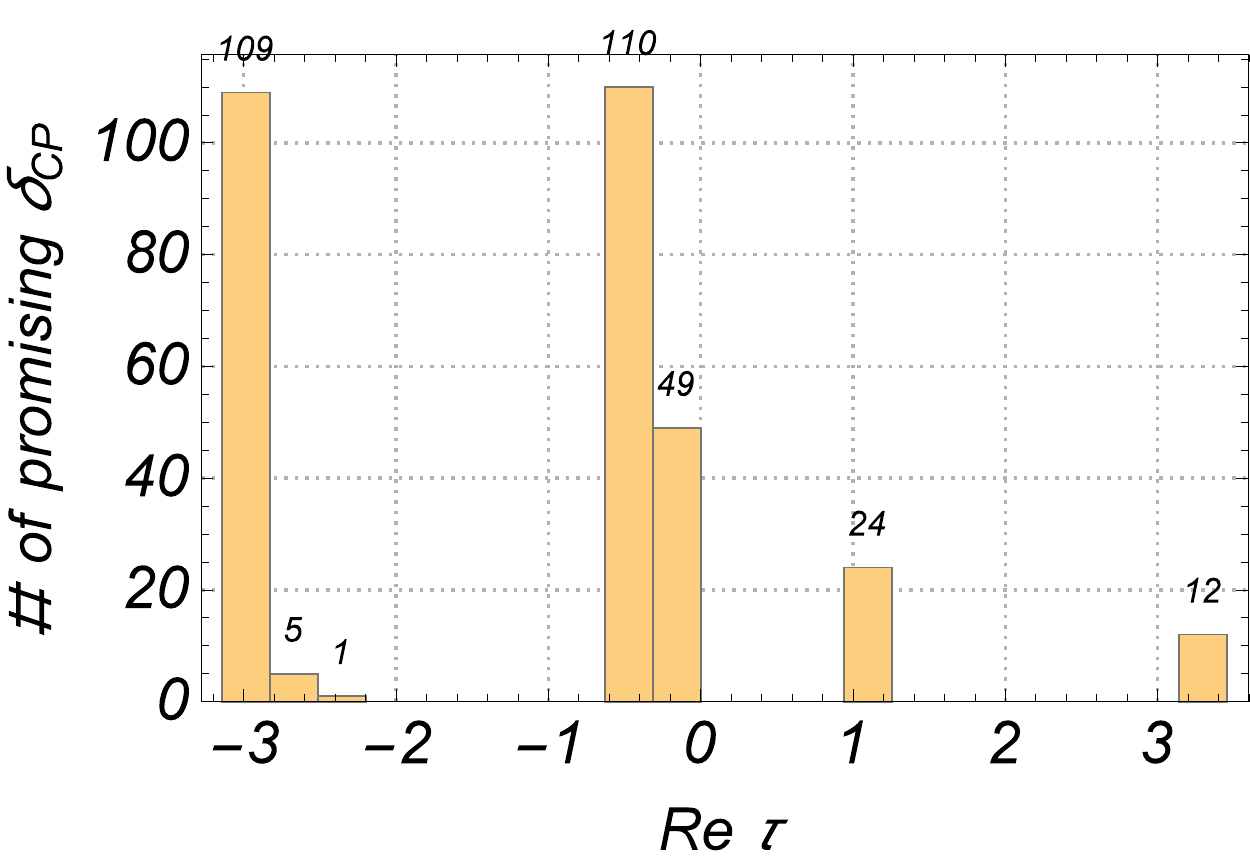}
\end{minipage}
\begin{minipage}{0.5\hsize}
\centering
\includegraphics[clip, width=0.85\hsize]{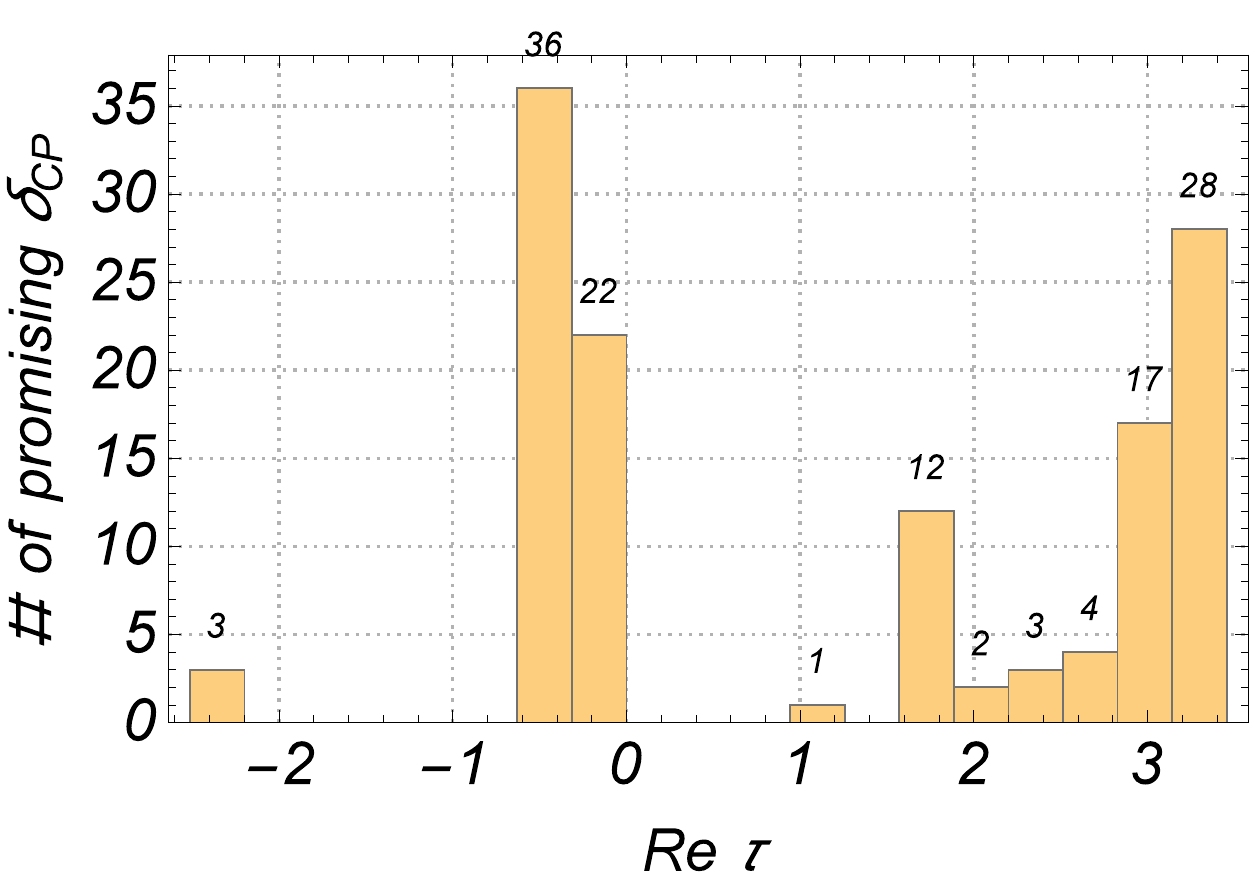}
\end{minipage}
\caption{Left panel: frequency of ${\rm Re}\, \tau$ satisfying an inequality in \eq{deltacpcut} for ${\rm Im} \, \tau=1.8$ in Pattern V.
{Digits on the top of} histogram bins denote the numbers of combinations of Higgs VEVs. Right panel: the same one for ${\rm Im} \, \tau=2.0$.}
\label{pattern5_3}
\end{figure}

\begin{figure}[H]
\begin{minipage}{0.5\hsize}
\centering
\includegraphics[clip, width=0.85\hsize]{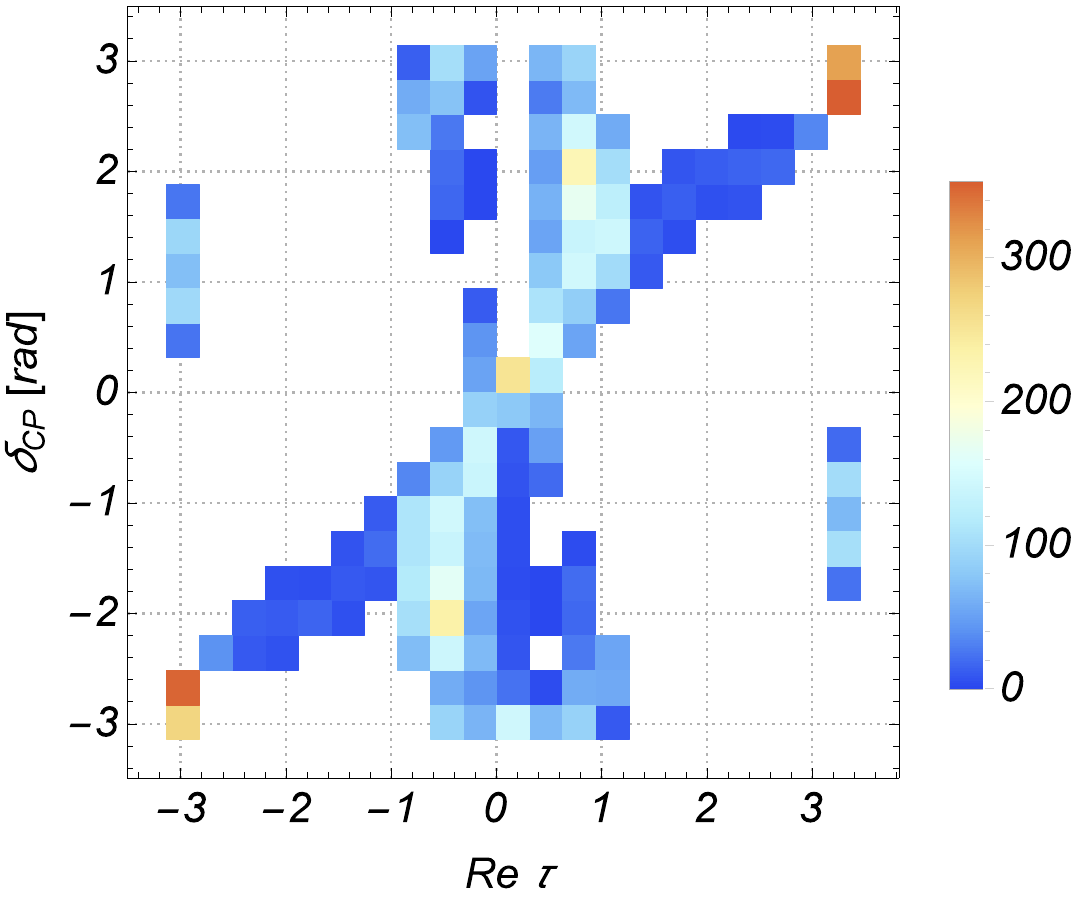}
\end{minipage}
\begin{minipage}{0.5\hsize}
\centering
\includegraphics[clip, width=0.85\hsize]{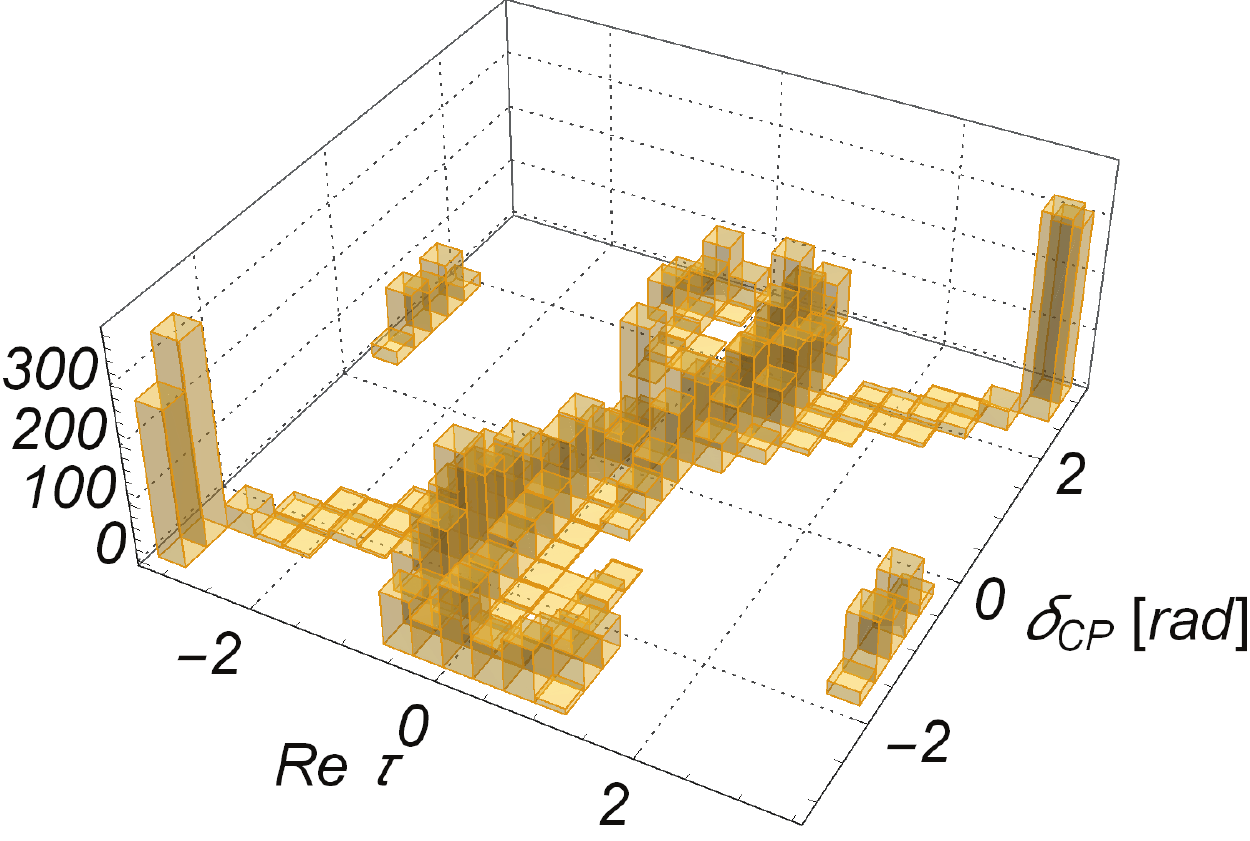}
\end{minipage}
\caption{Distributions of CP-violating phase $\delta_{\rm CP}$ vs. real part of complex structure modulus ${\rm Re}\, \tau$ for ${\rm Im} \, \tau=1.8$ in Pattern {\GkVII}.}
\label{pattern7_1}
\end{figure}

\begin{figure}[H]
\begin{minipage}{0.5\hsize}
\centering
\includegraphics[clip, width=0.85\hsize]{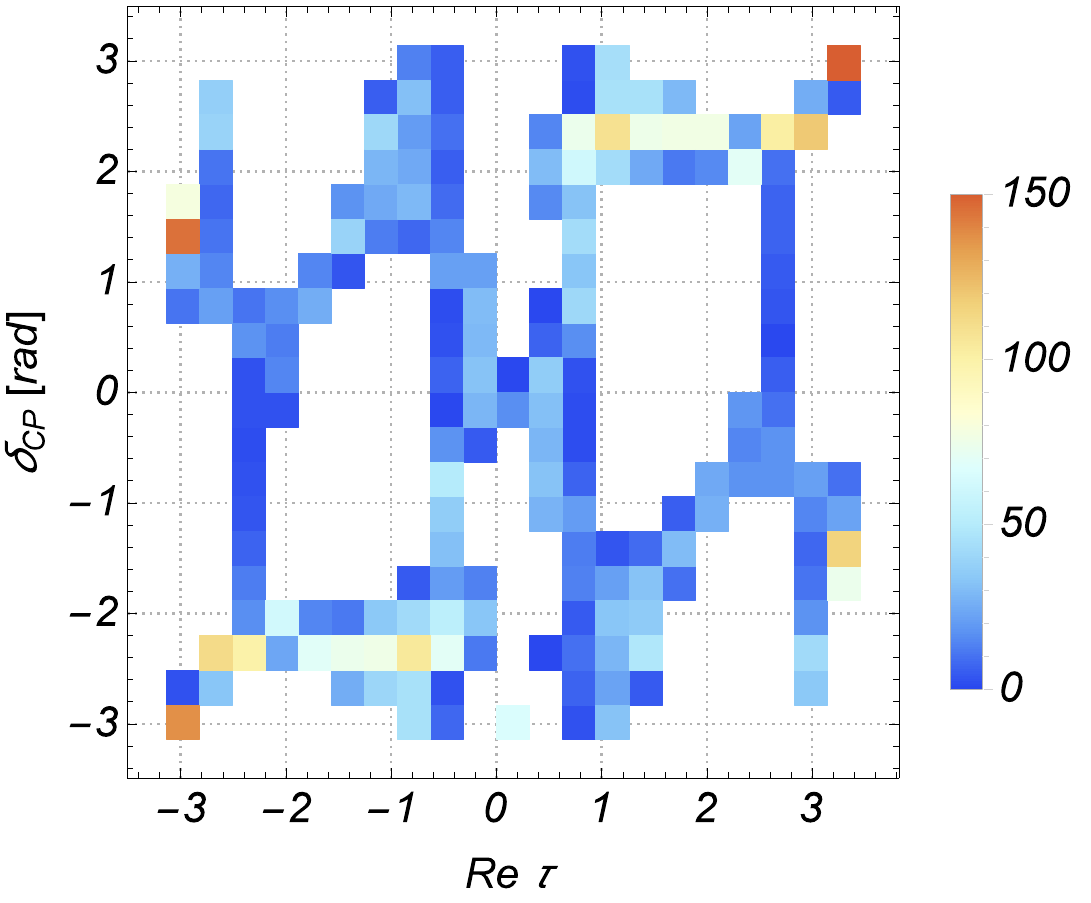}
\end{minipage}
\begin{minipage}{0.5\hsize}
\centering
\includegraphics[clip, width=0.85\hsize]{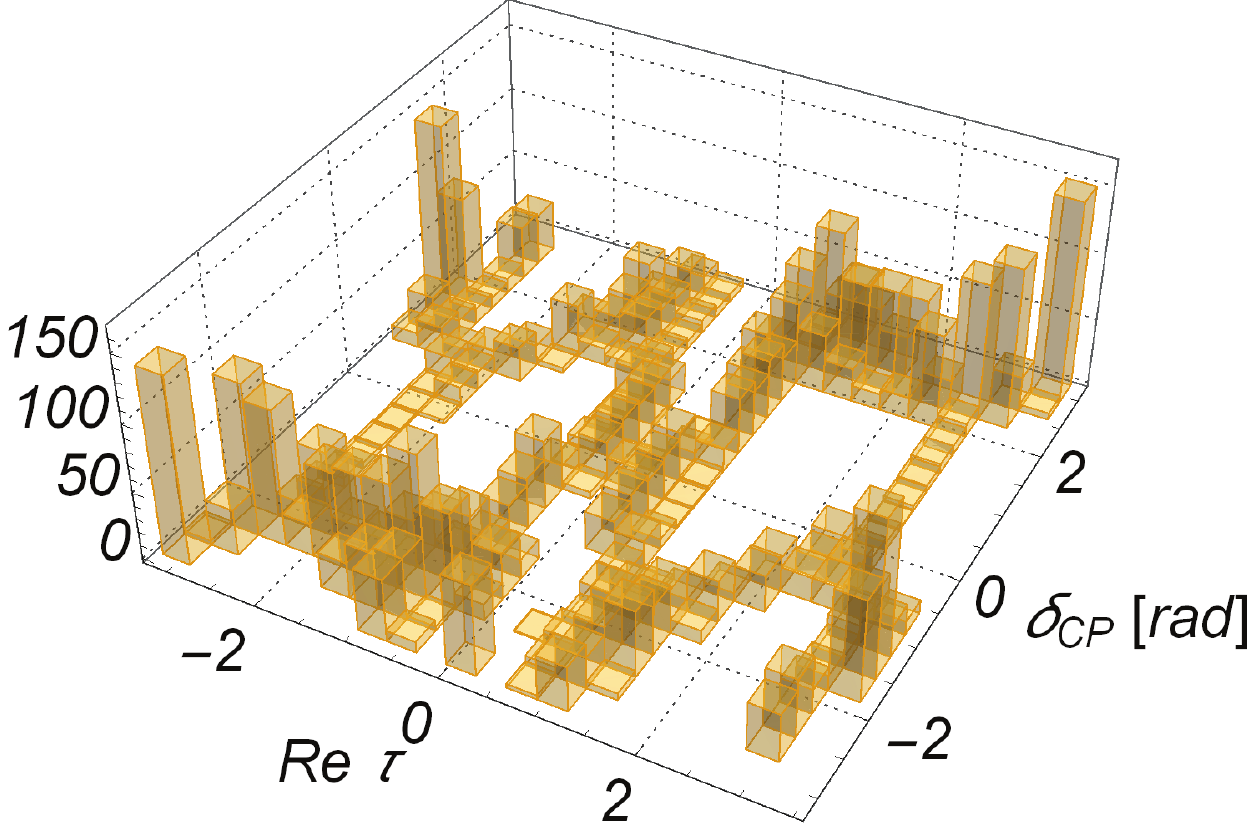}
\end{minipage}
\caption{Distributions of CP-violating phase $\delta_{\rm CP}$ vs. real part of complex structure modulus ${\rm Re}\, \tau$ for ${\rm Im} \, \tau=2.0$ in Pattern {\GkVII}.}
\label{pattern7_2}
\end{figure}

\begin{figure}[H]
\begin{minipage}{0.5\hsize}
\centering
\includegraphics[clip, width=0.85\hsize]{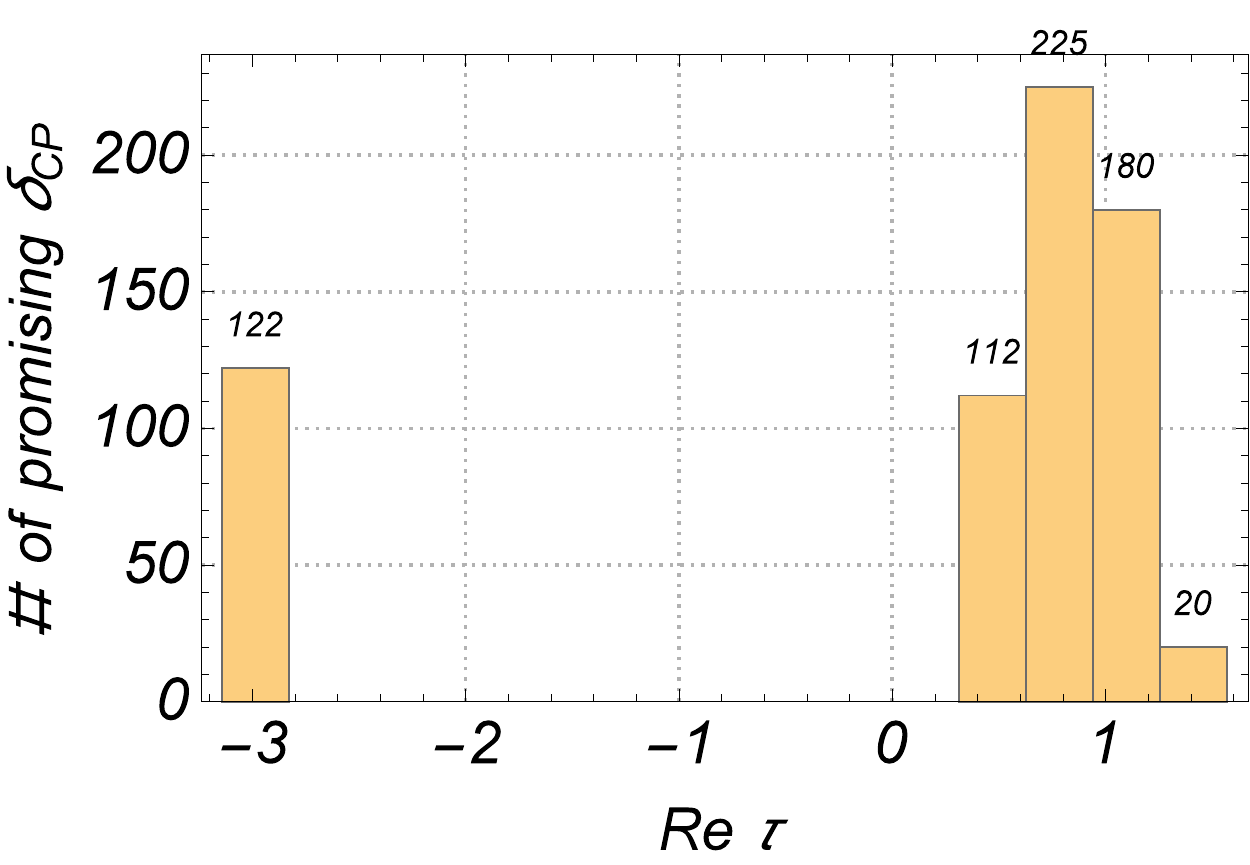}
\end{minipage}
\begin{minipage}{0.5\hsize}
\centering
\includegraphics[clip, width=0.85\hsize]{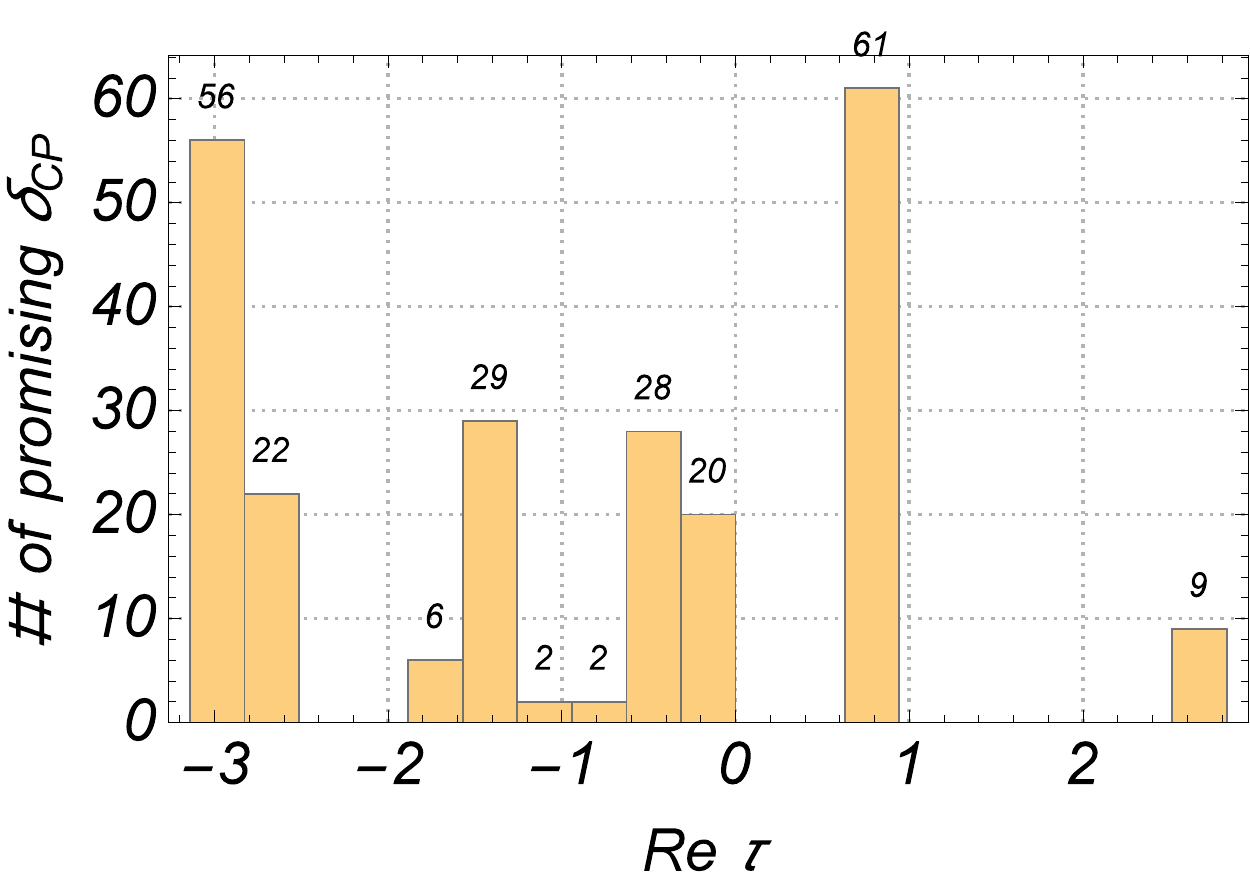}
\end{minipage}
\caption{Left panel: frequency of ${\rm Re}\, \tau$ satisfying an inequality in \eq{deltacpcut} for ${\rm Im} \, \tau=1.8$ in Pattern {\GkVII}.
{Digits on the top of} histogram bins denote the numbers of combinations of Higgs VEVs. Right panel: the same one for ${\rm Im} \, \tau=2.0$.}
\label{pattern7_3}
\end{figure}

The other samples of magnetic fluxes are shown in {Table}~\ref{fluxtable2}.
Here, three generations are realized in the quarks by taking $|M|=6$, where {nontrivial} SS twists are mandatory, $\alpha \neq 0$ and/or $\beta \neq 0$ \cite{Abe:2013bca, Abe:2015yva}.
Results of Pattern V are shown in {Figures}~\ref{pattern5_1}, \ref{pattern5_2} and \ref{pattern5_3}.
We could not find realistic flavor structures in Pattern {\GkVI}{,} where distributions of this pattern are not shown in the body.
{Figures}~\ref{pattern7_1}, \ref{pattern7_2} and \ref{pattern7_3} contain the results of Pattern {\GkVII}.
Different from the configuration of $M_{X}$ and $\eta_{X}$ in Eq.~(\ref{M_and_eta_configuration}), the most disfavored pattern {appears in the choice of the SS phases} $\alpha_{1,2} = 1/2$, $\beta_{1,2} = 0$ (Pattern {\GkVI}).

\begin{table}[H]
\centering
\begin{tabular}{|c|ccc|} \hline
 & $\{M_1, \alpha_1, \beta_1, \eta_1\}$ & $\{M_2, \alpha_2, \beta_2, \eta_2\}$ & $\{M_3, \alpha_3, \beta_3, \eta_3\}$ \\ \hline
Pattern V & $\{-6, 0, 1/2, +1\}$ & $\{-6, 0, 1/2, -1\}$ & $\{12, 0, 0, -1\}$ \\
Pattern {\GkVI} & $\{-6, 1/2, 0, +1\}$ & $\{-6, 1/2, 0, -1\}$ & $\{12, 0, 0, -1\}$ \\ 
Pattern {\GkVII} & $\{-6, 1/2, 1/2, +1\}$ & $\{-6, 1/2, 1/2, -1\}$ & $\{12, 0, 0, -1\}$ \\  \hline
\end{tabular}
\caption{The other samples of configurations.}
\label{fluxtable2}
\end{table}



%0
%\bibliographystyle{junsrt}

%\begin{multicols}{2}
%{\fontsize{10pt}{0pt}\selectfont
\bibliographystyle{JHEP}
\bibliography{references_JHEP_ver2}
%}
%\end{multicols}

%%%%%%%%%%%%%%%%%%%%%%%%%%%%%%%%%%%%%%%%%55
%\begin{thebibliography}{99}
%
%%\cite{Green:1987mn}
%\bibitem{Green:1987mn} 
%  M.~B.~Green, J.~H.~Schwarz and E.~Witten,
%  ``Superstring Theory. Vol. 2"%: Loop Amplitudes, Anomalies And Phenomenology,''
%  Cambridge, Uk: Univ. Pr. ( 1987) 596 P. ( Cambridge Monographs On Mathematical Physics)
%
%
%
%%\cite{Strominger:1985it}
%\bibitem{Strominger:1985it} 
%  A.~Strominger and E.~Witten,
%  %``New Manifolds for Superstring Compactification,''
%  Commun.\ Math.\ Phys.\  {\bf 101}, 341 (1985).
%%  doi:10.1007/BF01216094
%  %%CITATION = doi:10.1007/BF01216094;%%
% 
% %\cite{Dine:1992ya}
%\bibitem{Dine:1992ya} 
%  M.~Dine, R.~G.~Leigh and D.~A.~MacIntire,
%  %``Of CP and other gauge symmetries in string theory,''
%  Phys.\ Rev.\ Lett.\  {\bf 69}, 2030 (1992)
%%  doi:10.1103/PhysRevLett.69.2030
%  [hep-th/9205011].
%  %%CITATION = doi:10.1103/PhysRevLett.69.2030;%%
%  
% %\cite{Choi:1992xp}
%\bibitem{Choi:1992xp} 
%  K.~w.~Choi, D.~B.~Kaplan and A.~E.~Nelson,
%  %``Is CP a gauge symmetry?,''
%  Nucl.\ Phys.\ B {\bf 391}, 515 (1993)
%%  doi:10.1016/0550-3213(93)90082-Z
%  [hep-ph/9205202].
%  %%CITATION = doi:10.1016/0550-3213(93)90082-Z;%%
%
% %\cite{Kobayashi:2003gf}
%\bibitem{Kobayashi:2003gf} 
%  T.~Kobayashi and O.~Lebedev,
%  %``Heterotic string backgrounds and CP violation,''
%  Phys.\ Lett.\ B {\bf 565}, 193 (2003)
%%  doi:10.1016/S0370-2693(03)00751-2
%  [hep-th/0304212].
%  %%CITATION = doi:10.1016/S0370-2693(03)00751-2;%%
%  
%  
%\end{thebibliography}
%%%%%%%%%%%%%%%%%%%%%%%%%%%%%%%%%%%%%%%%%%%%%%%5  
\end{document}